\documentclass[%
 reprint,
superscriptaddress,
 amsmath,amssymb,
 aps,
floatfix,
showkeys
]{revtex4-2}

\usepackage{graphicx}
\usepackage{dcolumn}
\usepackage{bm}
\usepackage[utf8]{inputenc}
\usepackage{tabularx}
\usepackage[dvipsnames]{xcolor}
\usepackage{units}
\usepackage[euler]{textgreek}
\usepackage{upgreek}
\usepackage{derivative}
\usepackage{amsmath}
\usepackage{physics}
\usepackage{amssymb}
\usepackage[colorlinks=true,linkcolor=blue!50!black,urlcolor=blue!50!black,citecolor=blue!50!black]{hyperref}

\begin{document}

\preprint{APS}

\newcommand{\mpi}{\affiliation{Max-Planck-Institut f\"ur Physik, 80805 M\"unchen, Germany}}
\newcommand{\coimbra}{\affiliation{Also at: LIBPhys, Departamento de Fisica, Universidade de Coimbra, P3004 516 Coimbra, Portugal}}
\newcommand{\su}{\affiliation{Now at: The Oskar Klein Centre, Department of Physics, Stockholm University, AlbaNova 10691, Stockholm, Sweden}}

\mpi

\coimbra

\su

\author{A.~Bento}
  \mpi
  \coimbra 

\author{A.~Bertolini}
  \mpi

\author{L.~Canonica}
  \mpi 

\author{S.~Di~Lorenzo}
  \mpi

\author{F.~Dominsky}
    \mpi

\author{N.~Ferreiro~Iachellini}
  \mpi

\author{D.~Fuchs}
\email[Corresponding author: ]{dominik.fuchs@mpp.mpg.de}
  \mpi 
  \su

\author{A.~Garai}
  \mpi 
  
\author{D.~Hauff}
  \mpi 

\author{A.~Langenk\"amper}
  \mpi

\author{M.~Mancuso}
  \mpi
 
\author{B.~Mauri}
    \mpi
  
\author{F.~Petricca}
  \mpi 

\author{F.~Pr\"obst}
  \mpi

\author{F.~Pucci}
  \mpi

\author{L.~Stodolsky}
  \mpi

\noaffiliation

\title{Solar Neutrinos in Cryogenic Detectors}

\begin{abstract}
Coherent elastic neutrino-nucleus scattering (CE$\nu$NS) poses an irreducible background in the search for dark matter-nucleus elastic scatterings, which is commonly known as the neutrino floor. As direct dark matter search experiments keep improving their sensitivity into so far unexplored regions, they face the challenge of approaching this neutrino floor. A precise
description of the CE$\nu$NS signal is therefore crucial for the description of backgrounds
for future DM searches. In this work we discuss the scenario of detecting neutrinos in low-threshold, high-exposure cryogenic solid state experiments optimized for the search of low-mass dark matter. The energy range considered is completely dominated by solar neutrinos. In absence of any dark matter events, we treat solar neutrinos as the main signal of interest. We show that sensitivity to the flux of neutrinos from different production mechanisms can be achieved. In particular we investigate the sensitivity to the flux of pp and $^{7}$Be neutrinos, as well as CNO neutrinos. Furthermore, we investigate the sensitivity to dark matter signals in the presence of a solar neutrino background for different experimental scenarios, which are defined by three parameters: the target material, the energy threshold and the exposure. We show that experiments with thresholds of $\mathcal{O}$(eV) and exposures of $\mathcal{O}$(tonne-years), using CaWO$_{4}$ or Al$_{2}$O$_{3}$ targets, have discovery potential for dark matter interaction cross sections in the neutrino floor.
\end{abstract}

\maketitle



\section{Introduction}
A common challenge for the field of direct dark matter searches in the race towards unprecedented sensitivities is the approach of the so-called neutrino floor. These are events caused by coherent elastic neutrino-nucleus scattering (CE$\nu$NS) and are indistinguishable from dark matter events in the detector. The relevant neutrino sources typically considered in the calculation of the neutrino floor are solar neutrinos, atmospheric neutrinos and diffuse supernova neutrinos \cite{PhysRevD.89.023524}.

Massive liquid scintillator and liquid noble gas experiments with large exposures continue to improve their sensitivity for dark matter particle masses in the 10$\,$GeV/c$^{2}$ range. These experiments are already approaching levels at which the first neutrino-nucleus interactions, mostly with the high energetic solar neutrinos ($^{8}$B neutrinos), are expected (multi-tonne liquid xenon detectors \cite{Baudis_2014, Schumann_2015}, XENON1T experiment \cite{PhysRevLett.126.091301}, LZ experiment \cite{PhysRevD.101.052002}). Meanwhile, experiments using cryogenic solid state detectors are setting new standards in terms of energy thresholds (SuperCDMS \cite{PhysRevLett.127.061801, PhysRevD.105.112006}, CRESST \cite{PhysRevD.107.122003, Angloher2024, AngloherVUV}). In \cite{AngloherVUV}, an energy threshold of 6.7$\,$eV for nuclear recoils is reported, opening the parameter space below 100$\,$MeV/c$^{2}$ dark matter particle masses.

In the energy range considered in experiments that are optimized for low-mass dark matter searches, the expected neutrino flux is completely dominated by solar neutrinos. The flux of solar neutrinos increases by several orders of magnitude towards lower energies. This implies that the exposure at which the neutrino floor will become relevant is lower for experiments achieving very low energy thresholds.

In this paper we study the sensitivity to the solar neutrino fluxes in future experiments using cryogenic detectors for direct detection of dark matter with masses in the sub-GeV/c$^{2}$ regime. Being sensitive to signals of solar neutrinos is an opportunity to study certain aspects of the solar model, such as the solar metallicity and allow for solar model independent tests of the Mikheyev-Smirnov-Wolfenstein (MSW) effect. Not only do CE$\nu$NS have a much higher cross section compared to electron interactions, but they are also flavor independent and therefore reduce the uncertainties on the measured fluxes introduced by neutrino flavor conversion. A fully flavor independent measurement of the solar neutrino flux via neutral current interactions, so far not accessible at low energies, is also completely independent of uncertainties from the solar model. Therefore we explore the experimental conditions (exposures, thresholds and background levels) needed to have sensitivity to CE$\nu$NS of solar neutrinos, in particular for a measurement of pp, $^{7}$Be and CNO neutrinos, which have not been observed so far via CE$\nu$NS. 

Moreover, once these experiments reach the necessary exposures to become sensitive to CE$\nu$NS with solar neutrinos, having a precise description of the corresponding expected signals will also become absolutely crucial for the description of backgrounds for the dark matter search. Understanding the solar neutrino signal allows for having sensitivity to a potential dark matter signal even below the neutrino floor. Therefore, in this work, we also calculate discovery potentials for dark matter particles down to masses of 50$\,$MeV/c$^{2}$ in the presence of solar neutrinos, achievable in experiments with energy thresholds of $\mathcal{O}$(eV). Thresholds in this energy range are already reached by solid state experiments of the current generation \cite{AngloherVUV}. 

We start by giving a short introduction on the calculation of the expected recoil spectra of solar neutrinos and our likelihood framework in Sec. \ref{SpecAndL}. We then show the sensitivity to different fluxes of solar neutrinos in Sec. \ref{SolNu_Sens}. Afterwards, we treat solar neutrinos as a background for dark matter searches. We show the calculation of the neutrino floor and the calculation of sensitivity limits to dark matter signals in the presence of a solar neutrino background in Sec. \ref{DM_Sens}. At the end we summarize our results in Sec. \ref{conclusion}.

\section{Solar neutrino spectrum and likelihood framework} \label{SpecAndL}

In this section we briefly outline the calculation of the expected rate of solar neutrinos in a detector and introduce the likelihood function that we use in our analysis.

\subsection{Solar Neutrino recoil rate}

The neutrino rate is given by a convolution of the solar neutrino flux, $\Phi_{\nu}(E_{\nu})$, with the differential neutrino-nucleus scattering cross section: 

\begin{equation} \label{eq:SolNuRate}
	\dfrac{\mathrm{d}R_{\nu}}{\mathrm{d}E_{\mathrm{R}}} \propto \int\limits_{E_{\nu, \mathrm{min}}}^{\infty} dE_{\nu} \Phi_{\nu}(E_{\nu}) \dfrac{\mathrm{d}\sigma(E_{\mathrm{R}},E_{\nu})}{\mathrm{d}E_{\mathrm{R}}}
\end{equation}

The lower boundary of the integral over the neutrino energy $E_{\nu}$ is given by the minimal energy a neutrino needs to lead to a recoil with energy $E_{\mathrm{R}}$ in the target: $E_{\nu, \mathrm{min}}(E_{\mathrm{R}}) = \sqrt{m_{\mathrm{T}}E_{\mathrm{R}}/2}$, with $m_{\mathrm{T}}$ being the mass of the target nucleus.

The differential cross section is well understood within the standard model of particle physics. 
At low momentum transfers $q = \sqrt{2m_{\mathrm{T}}E_{\mathrm{R}}}$, 
the scattering is a coherent process. The cross section can be written as \cite{Cevns_article}:

\begin{equation} \label{eq:SolNuCrossSec}
	\dfrac{\mathrm{d}\sigma}{\mathrm{d}E_{\mathrm{R}}} = \dfrac{G_{F}^{2}}{2\pi} Q_{W}^{2} m_{\mathrm{T}}  \left(2 - \dfrac{E_{\mathrm{R}} m_{\mathrm{T}}}{E_{\nu}^{2}}\right) \, \vert F(q) \vert^{2}
\end{equation}

with $G_{F}$ being the Fermi constant and $Q_{W}$ being the weak nuclear charge, $Q_{W} \simeq \frac{1}{2} \left[ (4sin^{2}\theta_{W} - 1)Z + N \right]$, with $\theta_{W}$ being the Weinberg angle and Z and N being the number of protons and neutrons, respectively. We use the Helm form factor parameterization \cite{PhysRev.104.1466} for the nuclear form factor $\vert F(q) \vert^{2}$.

The total flux of solar neutrinos, $\Phi_{\nu}(E_{\nu})$, is the sum of the eight different individual fluxes of neutrino production mechanisms in the Sun. The exact flux of neutrinos produced in a specific fusion reaction depends on the choice of the solar model. Unless specified otherwise, all studies in this work are based on the model B16-GS98 \cite{Vinyoles_2017}. The corresponding values of the neutrino flux on Earth from each individual production mechanism are listed in Tab. \ref{tab:SolarModel}.

\begin{table*}[!htb]
\centering
	\caption{\label{tab:SolarModel}All neutrino production reactions and the corresponding neutrino fluxes on Earth according to the high-metallicity (HZ) model B16-GS98 and the low-metallicity (LZ) model B16-AGSS09met \cite{Vinyoles_2017}. The uncertainties represent the theoretical model errors and correlations among observables obtained from MC simulations in \cite{Vinyoles_2017}.}
	\newcolumntype{C}{>{\centering\arraybackslash}X}
	\setlength\extrarowheight{3pt}
	\noindent
    \begin{tabularx}{\textwidth}{ C c C C}
    \hline
    Solar & Production & B16-GS98 (HZ) & B16-AGSS09met (LZ) \\
    neutrinos & mechanism & flux $\phi$ (cm$^{-2}\,$s$^{-1}$) & flux $\phi$ (cm$^{-2}\,$s$^{-1}$) \\ \hline
    pp & p + p $\rightarrow$ $^2$H + e$^+$ + $\nu_{e}$ & 5.98$\,$(1$\,\pm\,$0.006)$\,\cdot\,$10$^{10}$ & 6.03$\,$(1$\,\pm\,$0.005)$\,\cdot\,$10$^{10}$ \\
    pep & p + e$^-$ + p $\rightarrow$ $^2$H + $\nu_{e}$ & 1.44$\,$(1$\,\pm\,$0.01)$\,\cdot\,$10$^8$ & 1.46$\,$(1$\,\pm\,$0.009)$\,\cdot\,$10$^8$ \\
    hep & $^3$He + p $\rightarrow$ $^4$He + e$^+$ + $\nu_{e}$ & 7.98$\,$(1$\,\pm\,$0.30)$\,\cdot\,$10$^3$ & 8.25$\,$(1$\,\pm\,$0.30)$\,\cdot\,$10$^3$ \\
    $^{7}$Be & $^{7}$Be + e$^-$ $\rightarrow$ $^{7}$Li + $\nu_{e}$ & 4.93$\,$(1$\,\pm\,$0.06)$\,\cdot\,$10$^9$ & 4.50$\,$(1$\,\pm\,$0.06)$\,\cdot\,$10$^9$ \\
    $^8$B & $^8$B $\rightarrow$ $^8$Be$^*$ + e$^+$ + $\nu_{e}$ & 5.46$\,$(1$\,\pm\,$0.12)$\,\cdot\,$10$^6$ & 4.50$\,$(1$\,\pm\,$0.12)$\,\cdot\,$10$^6$ \\
    $^{13}$N & $^{13}$N $\rightarrow$ $^{13}$C + e$^+$ + $\nu_{e}$ & 2.78$\,$(1$\,\pm\,$0.15)$\,\cdot\,$10$^8$ & 2.04$\,$(1$\,\pm\,$0.14)$\,\cdot\,$10$^8$ \\
    $^{15}$O & $^{15}$O $\rightarrow$ $^{15}$N + e$^+$ + $\nu_{e}$ & 2.05$\,$(1$\,\pm\,$0.17)$\,\cdot\,$10$^8$ & 1.44$\,$(1$\,\pm\,$0.16)$\,\cdot\,$10$^8$ \\ 
	$^{17}$F & $^{17}$F $\rightarrow$ $^{17}$O + e$^+$ + $\nu_{e}$ & 5.29$\,$(1$\,\pm\,$0.20)$\,\cdot\,$10$^6$ & 3.26$\,$(1$\,\pm\,$0.18)$\,\cdot\,$10$^6$ \\ \hline
    \end{tabularx}
\end{table*}

\begin{figure}[h]
	\centering
	\includegraphics[width=0.48\textwidth]{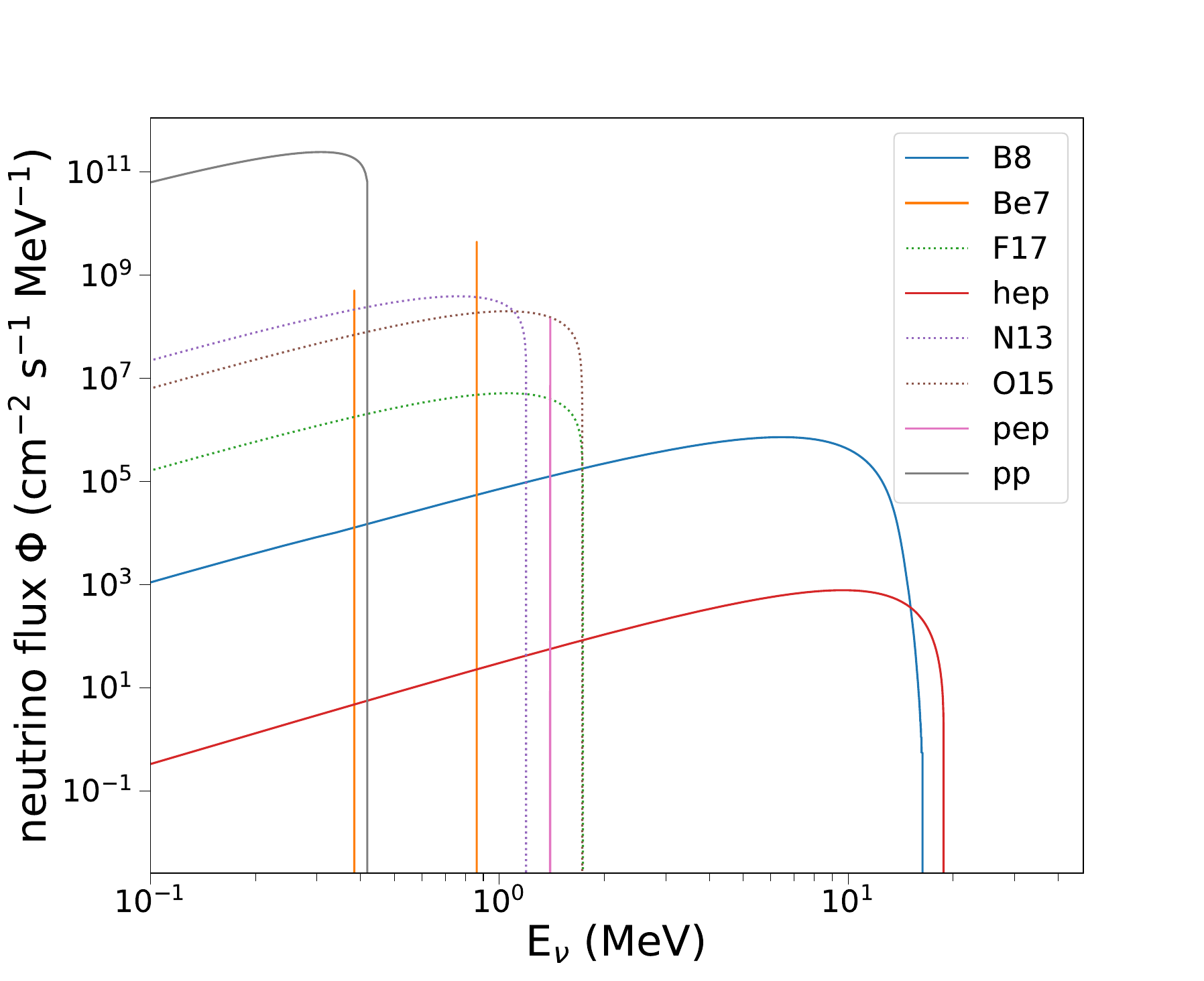}
	\caption{\label{fig:FluxSpectra}Solar neutrino energy spectra. Neutrinos produced in the fusion reactions of the pp-chain are shown with solid lines, neutrinos produced in the CNO-cycle with dashed lines.}
\end{figure}

While the flux of the neutrinos from different reactions is given by the solar model, the shape of their energy spectra just depends on the fusion reactions in which they are produced. The solar neutrino energy spectra are shown in Fig. \ref{fig:FluxSpectra}, scaled with the corresponding fluxes of Tab. \ref{tab:SolarModel}.

The studies presented in this work can be done for a variety of different target materials that are typically used in the field of DM direct detection and CE$\nu$NS experiments. The advantage of composite materials like CaWO$_{4}$ is the presence of both, heavy and light nuclei. Heavy nuclei have a large cross section for CE$\nu$NS due to the approximate relation of $Q_{W} \propto N$ in eq. \ref{eq:SolNuCrossSec}, while light nuclei help to extend the sensitivity to low neutrino energies due to kinematics. The advantage of using different materials is the difference in the shape of their recoil spectra, which can be used to confirm potential signals among different detectors. We choose CaWO$_{4}$ and Al$_{2}$O$_{3}$ as target materials, which are used e.g. by CRESST \cite{PhysRevD.100.102002, AngloherVUV} and NUCLEUS \cite{Rothe2020}.

\begin{figure}[h]
	\centering
		\includegraphics[width=0.48\textwidth]{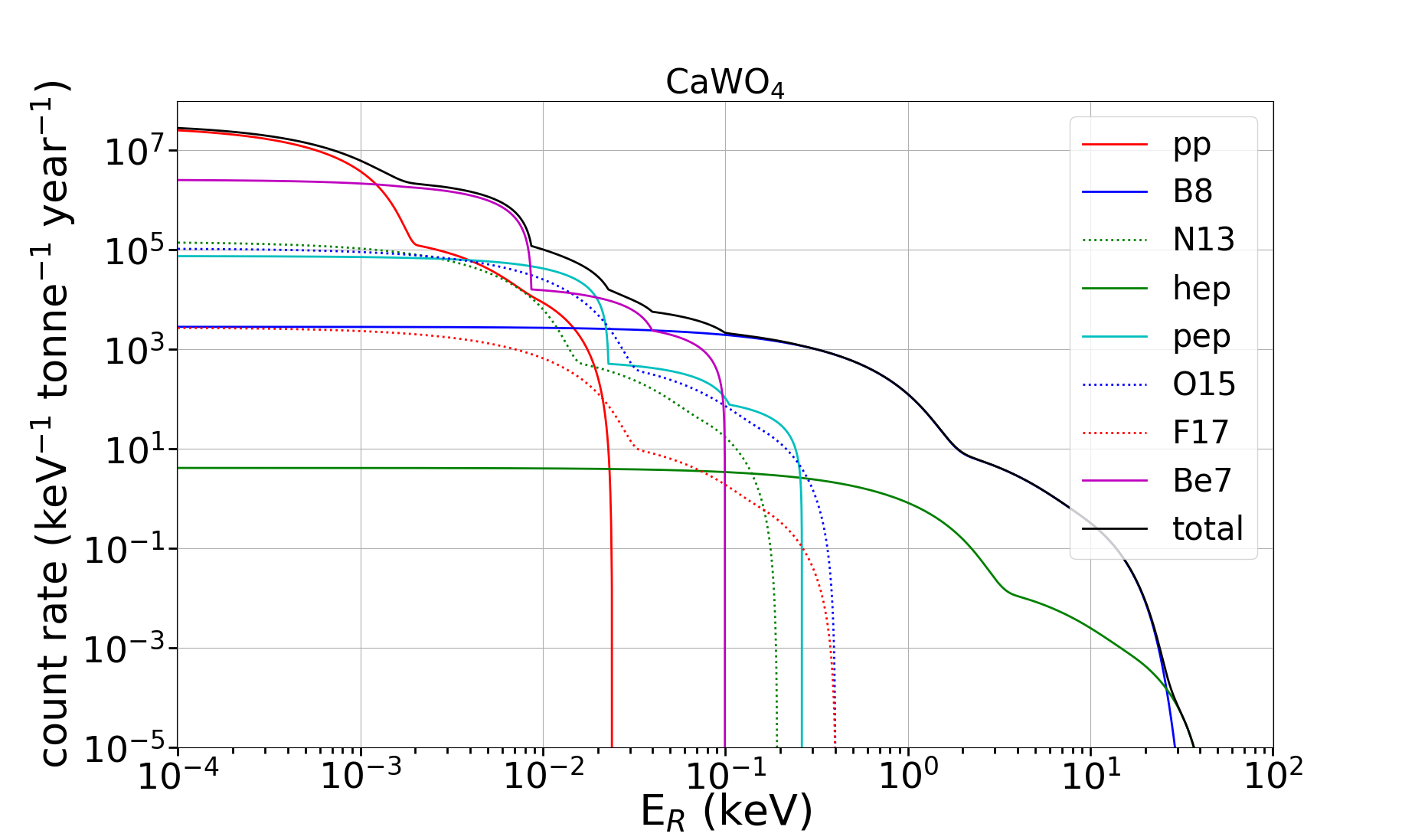}
		\caption{\label{fig:SolNuRecoilSpec}Differential recoil spectrum of solar neutrinos in a CaWO$_{4}$ detector.}
\end{figure}

\begin{figure}[h]
	\centering
		\includegraphics[width=0.48\textwidth]{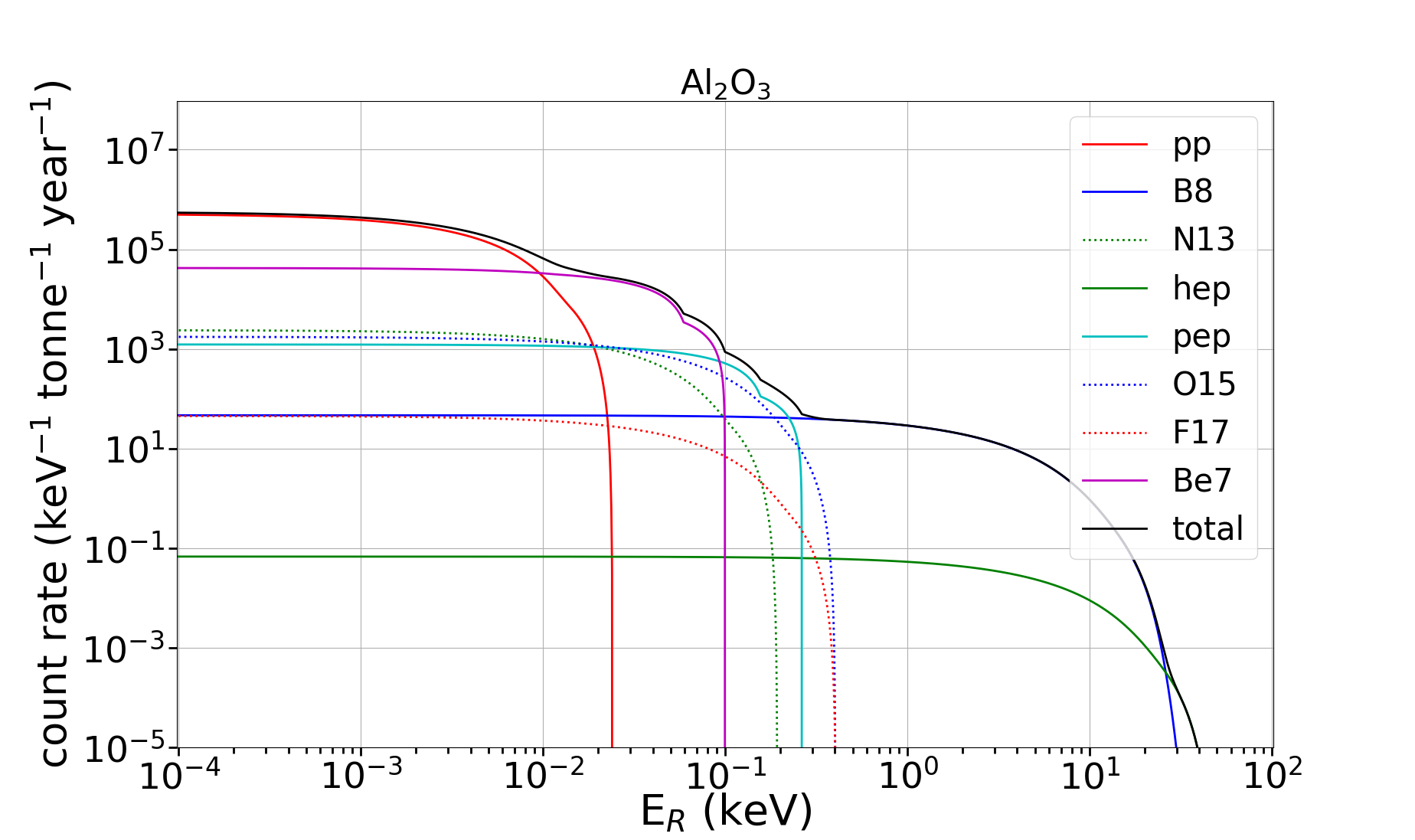}
		\caption{\label{fig:SolNuRecoilSpec_Sapp}Differential recoil spectrum of solar neutrinos in a Al$_{2}$O$_{3}$ detector.}
\end{figure}

The expected energy spectra calculated with eq. \ref{eq:SolNuRate} for a CaWO$_{4}$ target is shown in Fig. \ref{fig:SolNuRecoilSpec} and for a Al$_{2}$O$_{3}$ target in Fig. \ref{fig:SolNuRecoilSpec_Sapp}.

Because of the larger cross section with heavier elements, the event rate is generally higher in CaWO$_{4}$ due to the scatterings on W and Ca, which are heavier than Al and O. On the other hand, scatterings on Al and O extend to higher recoil energies, due to kinematics. The lightest element (O) is the same in both crystals. Therefore the endpoint of all different neutrino spectra are the same in both materials, with the highest recoil energies being at $\mathcal{O}$(10$\,$keV), caused by scatterings of $^{8}$B and $hep$ neutrinos. Al$_{2}$O$_{3}$ crystals have a higher oxygen content (in comparison to CaWO$_{4}$ crystals of the same mass), leading to a higher rate in the high energy tail of the individual neutrino spectra. This can be seen e.g. in the $^{8}$B recoil spectrum. The maximum recoil energy of scatterings of $^{8}$B neutrinos on Ca is at $\sim$2$\,$keV. Only scatterings on O extend to energies above this value. Due to its higher oxygen content the recoil rate is larger in Al$_{2}$O$_{3}$ above this energy. This also can be seen in the rate of pp neutrinos: In Al$_{2}$O$_{3}$, scatterings of pp neutrinos are dominant in an energy range above the threshold until $\sim$10$\,$eV, while in CaWO$_{4}$ the rate is dominant in the energy range until $\sim$2$\,$eV. Unless an experiment achieves sub-eV energy thresholds, this makes Al$_{2}$O$_{3}$ detectors more suitable for the search of pp neutrinos.

The total number of expected events as a function of the energy threshold, $E_{\mathrm{thr}}$, is given by an integral over the recoil rate:

\begin{equation}
	R_{\nu}(E_{\mathrm{thr}}) = \int\limits_{E_{\mathrm{thr}}}^{\infty} \dfrac{\mathrm{d}R_{\nu}}{\mathrm{d}E_{\mathrm{R}}} \, \mathrm{d}E_{\mathrm{R}}
\end{equation}

\subsection{Likelihood framework}
All results of this paper are obtained with extended maximum likelihood fits, based on a binned poissonian likelihood function, following the approach in \cite{PhysRevD.90.083510}, defined as:

\begin{equation} \label{eq:BinnedL}
	\mathcal{L} = \prod\limits_{i=0}^{m} e^{-N_{e} f_{i}} \cdot \dfrac{(N_{e}f_{i})^{k_{i}}}{k_{i}!} \cdot \prod\limits_{j=1}^{8} \mathcal{L}_{\nu,j}(\Phi_{\nu,j})
\end{equation}

with $m$ being the number of bins and $k_{i}$ being the observed number of events in the $i$th bin. $N_{e}$ is the total number of expected events and $f_{i}$ is the binned probability density function (for the $i$th bin), defined as:

\begin{equation} \label{eq:PDFNorm}
	f_{i} = \dfrac{1}{N_{e}} \cdot \int\limits_{E_{\mathrm{R},i}^{\mathrm{low}}}^{E_{\mathrm{R},i}^{\mathrm{up}}} \tilde{f_{\mathrm{b}}}(E_{\mathrm{R}},r_{\mathrm{b}}) + \sum\limits_{\nu} \tilde{f_{\nu}}(E_{\mathrm{R}},\phi_{\nu}) \mathrm{d}E_{\mathrm{R}}
\end{equation}

with:

\begin{equation} \label{eq:expected}
	N_{e} = \int\limits_{E_{\mathrm{thr}}}^{\infty} \tilde{f_{\mathrm{b}}}(E_{\mathrm{R}},r_{\mathrm{b}}) + \sum\limits_{\nu} \tilde{f_{\nu}}(E_{\mathrm{R}},\phi_{\nu}) \mathrm{d}E_{\mathrm{R}}
\end{equation}

The function $\Tilde{f_{\mathrm{b}}}$ is describing a flat background with rate $r_{\mathrm{b}}$ and the function $\Tilde{f_{\nu}}$ corresponds to the solar neutrinos, respectively scaled to the exposure, $\epsilon$, which is the product of the total detector mass and measuring time. The fit parameters are the background rate $r_{\mathrm{b}}$ and the flux normalization of the eight neutrino fluxes, $\phi_{\nu}$. The recoil rate of the neutrinos is calculated with the flux normalizations of the theoretical model parameters, $\phi_{\nu,M}$, of Tab. \ref{tab:SolarModel}, making the free parameters $\phi_{\nu}$ simply a scaling factor of the theoretical expectation.

\begin{equation} \label{eq:UnNormPDF}
\begin{split}
	\tilde{f_{\mathrm{b}}}(E_{\mathrm{R}},r_{\mathrm{b}}) = \epsilon \cdot r_{\mathrm{b}} \cdot \dfrac{\mathrm{d}R_{\mathrm{b}}}{\mathrm{d}E_{\mathrm{R}}}(E_{\mathrm{R}}) \qquad \qquad \quad \, \\
	\tilde{f_{\nu}}(E_{\mathrm{R}},\phi_{\nu}) = \epsilon \cdot \phi_{\nu} \cdot \dfrac{\mathrm{d}R_{\nu}}{\mathrm{d}E_{\mathrm{R}}}(E_{\mathrm{R}},\phi_{\nu} = \phi_{\nu,M})
\end{split}
\end{equation}

with $\frac{\mathrm{d}R_{\mathrm{b}}}{\mathrm{d}E_{\mathrm{R}}}$ being a normalized uniform flat distribution.

The last terms of eq. \ref{eq:BinnedL}, $\mathcal{L}_{\nu,j}(\Phi_{\nu,j})$, are pull-terms parameterized as gaussian distributions with means at the theoretical expectations, $\phi_{\nu,M}$, and standard deviations corresponding to the theoretical uncertainties, $\sigma_{\nu}$ in the last column of Tab. \ref{tab:SolarModel}. These pull-terms can be individually included or excluded in the likelihood, giving constraints to the flux parameters of a chosen subset of neutrino fluxes.

\section{Sensitivity to Solar Neutrinos}\label{SolNu_Sens}

A fully flavor independent measurement of the solar flux has the potential of providing new strong constraints on neutrino physics and on the solar model. The purely neutral current (NC) measurement of solar neutrinos, so far not accessible at low energies, combined with the electron scatterings (NC + CC) of the solar neutrino spectrum, performed e.g. by Borexino \cite{Agostini2018}, SNO \cite{PhysRevC.88.025501} and Superkamiokande \cite{PhysRevD.94.052010}, would allow tests of the MSW effect \cite{smirnov2003msw}. Deviations from the expectations of the MSW effect could be a hint towards physics beyond the standard model, like e.g. the existence of sterile neutrinos \cite{smirnov2003msw, BOSER2020103736}. A precise measurement of the CNO neutrinos could shed light on the "solar metallicity problem" \cite{Cerdeño_2018}. Therefore, in this section we test the possibility of measuring solar neutrinos via CE$\nu$NS and the sensitivity to the different neutrino fluxes. \\

In the following we will adress two different questions: \\

1) \textbf{How well can the experiment reconstruct the fluxes of the dominant low energy neutrino components ($^{7 }$Be and pp)?} For this question, we constrain all sub dominant components (all CNO neutrinos and $hep$ neutrinos) via the pull-terms in the likelihood function (around their HZ model values of Tab. \ref{tab:SolarModel}), while all other neutrino components and the background rate are left as fully free parameters in the fit. We investigate which experimental settings need to be achieved to reach uncertainties below the theoretical ones in Tab. \ref{tab:SolarModel}. \\

2) \textbf{How sensitive is the experiment to the metallicity of the Sun?} For this question, we constrain all neutrinos of the pp-chain (with exception of the $^{7}$Be flux) via the pull-terms and the values of Tab. \ref{tab:SolarModel}, respectively using the HZ or LZ models, while the fluxes of neutrinos from the CNO-cycle and $^{7}$Be neutrinos are left as free parameters.

\subsection{Reconstruction of $^{7}$Be and pp neutrinos}

The neutrino flux parameters of $^{13}$N, $^{15}$O, $^{17}$F and hep neutrinos are constrained in the fit with pull-terms around their model value and within their model uncertainty, assuming the HZ solar model of Tab. \ref{tab:SolarModel}. The reconstruction of $^{7}$Be neutrinos is tested with CaWO$_{4}$ detectors for different experimental settings, defined by the energy threshold, exposure and flat background level. For a given combination of these experimental parameters, we generate 1000 Monte Carlo datasets based on the HZ model. We fit each dataset with the previously described likelihood function. Then we extract the mean and standard deviation of all reconstructed values of the $^{7}$Be flux parameter and compare it to the theoretical uncertainty of 6$\,\%$ (Tab. \ref{tab:SolarModel}) and the currently best measurement of the $^{7}$Be flux and its uncertainties 4.99$\,$(1$\,^{+0.025}_{-0.027}$)$\,\cdot\,$10$^{9}\,$cm$^{-2}\,$s$^{-1}$ \cite{Agostini2018}.

\begin{figure}[htb!]
	\centering
	\includegraphics[width=0.48\textwidth]{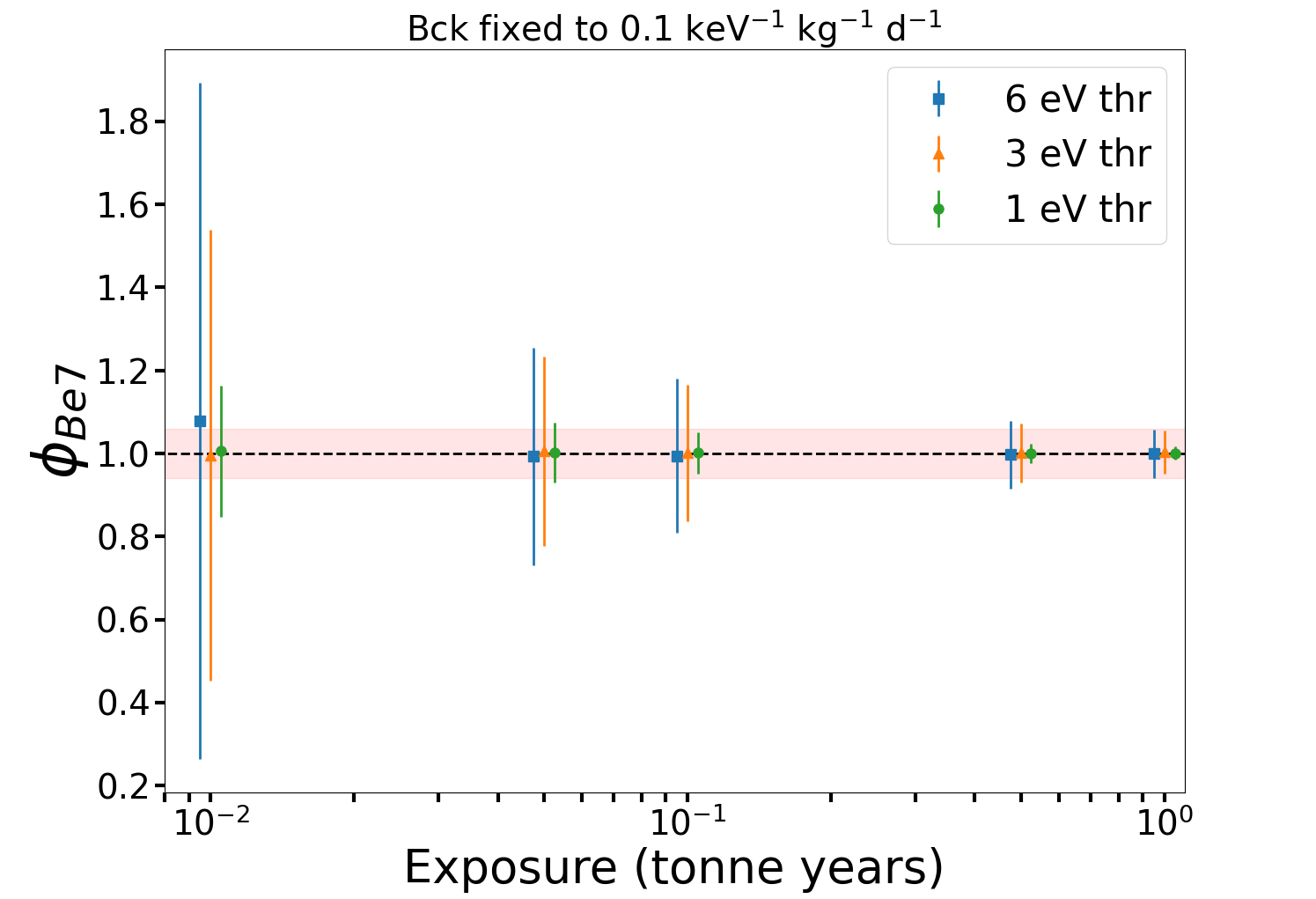}
	\caption{\label{fig:Be7_ExpoScan} The data points and error bars represent the mean and standard deviation of the reconstructed $^{7}$Be neutrino flux normalization parameter of 1000 Monte Carlo simulations with different experimental settings. The background rate is fixed at 0.1/(keV$\,$kg$\,$d) for all settings. The exposure is increased from 0.01$\,$tonne$\,$years to 1$\,$tonne$\,$year for three different thresholds, indicated by the legend. The data points are slightly shifted around the exposure of the simulations for visibility. The dashed line shows the model input value of the $^{7}$Be flux and the red shaded band represents the theoretical uncertainty.}
\end{figure}

In a first step, we fix the flat background level at a low value of 0.1$\,$counts/(keV$\,$kg$\,$d). In this nearly background free setting we then compare the accuracy of reconstructing the $^{7}$Be flux to the theoretical uncertainty as a function of the exposure for three different thresholds (1$\,$eV, 3$\,$eV, 6$\,$eV), which are chosen to be in the $\mathcal{O}$(eV) regime, which is close to what is already achievable in current experiments \cite{AngloherVUV}. The results are shown in Fig. \ref{fig:Be7_ExpoScan}

\begin{figure}[htb!]
	\centering
	\includegraphics[width=0.48\textwidth]{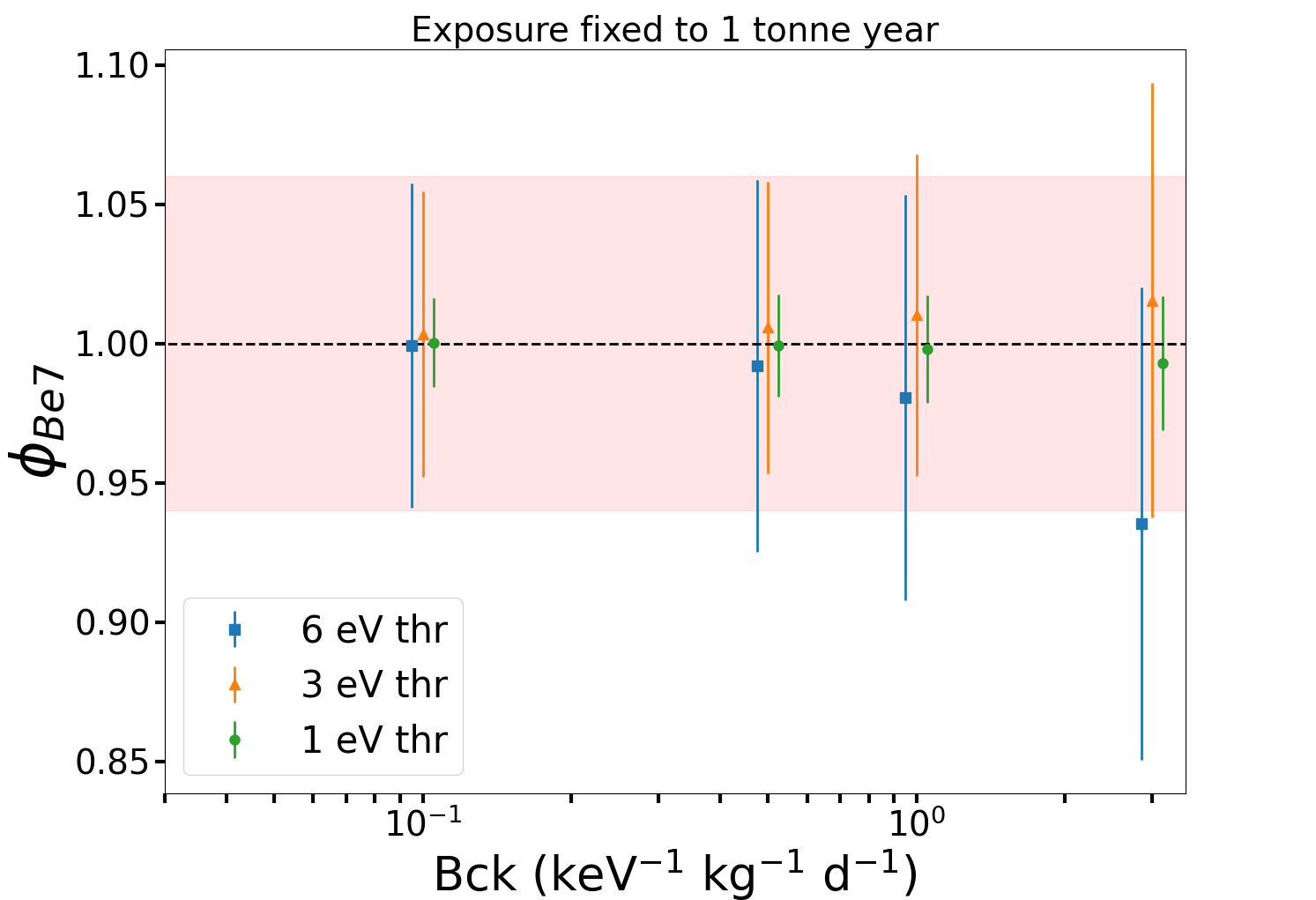}
	\caption{\label{fig:Be7_BckScan} The data points and error bars represent the mean and standard deviation of the reconstructed $^{7}$Be neutrino flux normalization parameter of 1000 Monte Carlo simulations with different experimental settings. The exposure is fixed at 1$\,$tonne$\,$years for all settings. The background rate is increased from 0.1/(keV$\,$kg$\,$d) to 3/(keV$\,$kg$\,$d) for three different thresholds, indicated by the legend. The data points are slightly shifted around the background level of the simulations for visibility. The dashed line shows the model input value of the $^{7}$Be flux and the red shaded band represents the theoretical uncertainty.}
\end{figure}

The results show that the flux is correctly reconstructed in all cases with exposures above 0.01$\,$tonne$\,$years. With a threshold of 1$\,$eV, an exposure of 0.1$\,$tonne$\,$years is enough to reach an uncertainty of 5.0$\,\%$, which is below the theoretical value of 6$\,\%$. With thresholds of 3$\,$eV or 6$\,$eV, the uncertainties (5.1$\,\%$ and 5.8$\,\%$, respectively) reach the benchmark at an exposure of 1$\,$tonne$\,$years and above. For a threshold of 1$\,$eV and an exposure of 1$\,$tonne$\,$years we additionally investigated the effect of constraining the fluxes of CNO and \textit{hep} neutrinos to the LZ model values, in which case the mean of the reconstructed $^{7}$Be flux shows a small bias of 1.5$\,\%$ above the injected value to compensate for the lower CNO rates. In a second step, we fix the exposure at 1$\,$tonne$\,$years. Analogous to before, we compare the accuracy of the reconstruction of the $^{7}$Be flux to the theoretical uncertainty, now as a function of the background rate. We start at the low value of 0.1$\,$counts/(keV$\,$kg$\,$d), which can be approximately considered background free, and increase it towards values in the $\mathcal{O}$(1$\,$counts/(keV$\,$kg$\,$d)) regime. The lowest flat background levels achieved in the CRESST experiment were at 5.1$\,$counts/(keV$\,$kg$\,$d) in an energy range of 1$\,$-$\,$16$\,$keV \cite{PhysRevD.100.102002} and 3.51$\,$counts/(keV$\,$kg$\,$d) in an energy range of 1$\,$-$\,$40$\,$keV \cite{Strauss_2015}. We scan up to a value of 3$\,$counts/(keV$\,$kg$\,$d). The results are shown in Fig. \ref{fig:Be7_BckScan}. \\

\begin{figure}[htb!]
	\centering
	\includegraphics[width=0.48\textwidth]{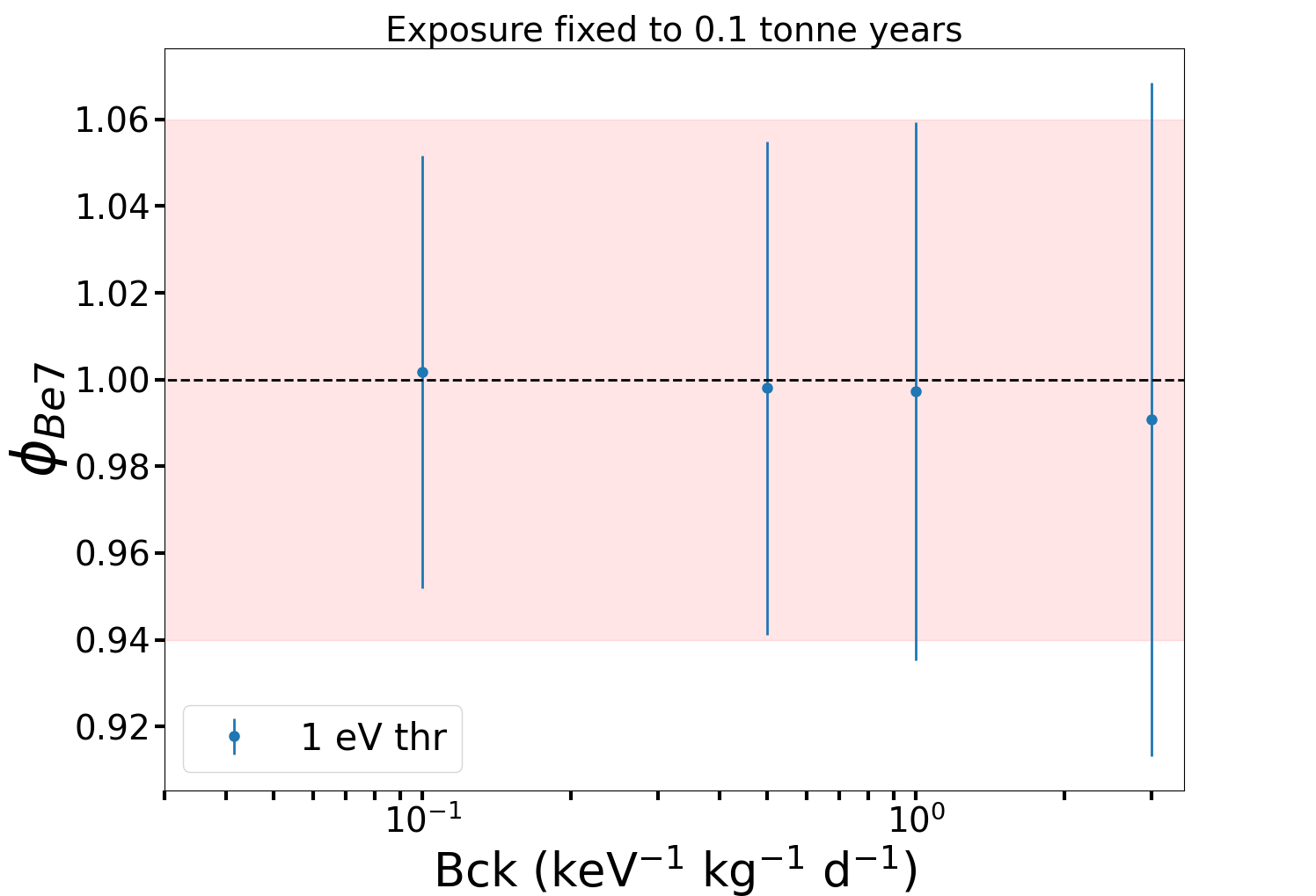}
	\caption{\label{fig:Be7_BckScan1eV} The data points and error bars represent the mean and standard deviation of the reconstructed $^{7}$Be neutrino flux normalization parameter of 1000 Monte Carlo simulations. The exposure is fixed at 0.1$\,$tonne$\,$years and the threshold is fixed at 1$\,$eV for all points. The background rate is increased from 0.1/(keV$\,$kg$\,$d) to 3/(keV$\,$kg$\,$d). The dashed line shows the model input value of the $^{7}$Be flux and the red shaded band represents the theoretical uncertainty.}
\end{figure}

Around the background level of about 1$\,$count/(keV$\,$kg$\,$d), a small bias in the reconstructed fluxes becomes apparent for the higher thresholds, which increases at higher background levels. At a background level above 1$\,$count/(keV$\,$kg$\,$d) the uncertainties on the reconstruction of the $^{7}$Be flux reach or exceed the theoretical benchmark for the thresholds of 3$\,$eV and 6$\,$eV (5.8$\,\%$ and 7.3$\,\%$, respectively). The reconstructed values of the $^{7}$Be flux for the energy threshold of 1$\,$eV show no deviation from the injected value and the uncertainties stay well below the benchmark. At the largest background level of 3$\,$counts/(keV$\,$kg$\,$d), the uncertainty is at 2.4$\,\%$, which is slightly below the currently best experimental uncertainty of $^{+2.5}_{-2.7}\,\%$ \cite{Agostini2018}. We also perform the same study for a setup with a fixed exposure of 0.1$\,$tonne$\,$years and a threshold of 1$\,$eV. The results can be seen in \mbox{Fig. \ref{fig:Be7_BckScan1eV}}.

\begin{figure}[htb!]
	\centering
	\includegraphics[width=0.48\textwidth]{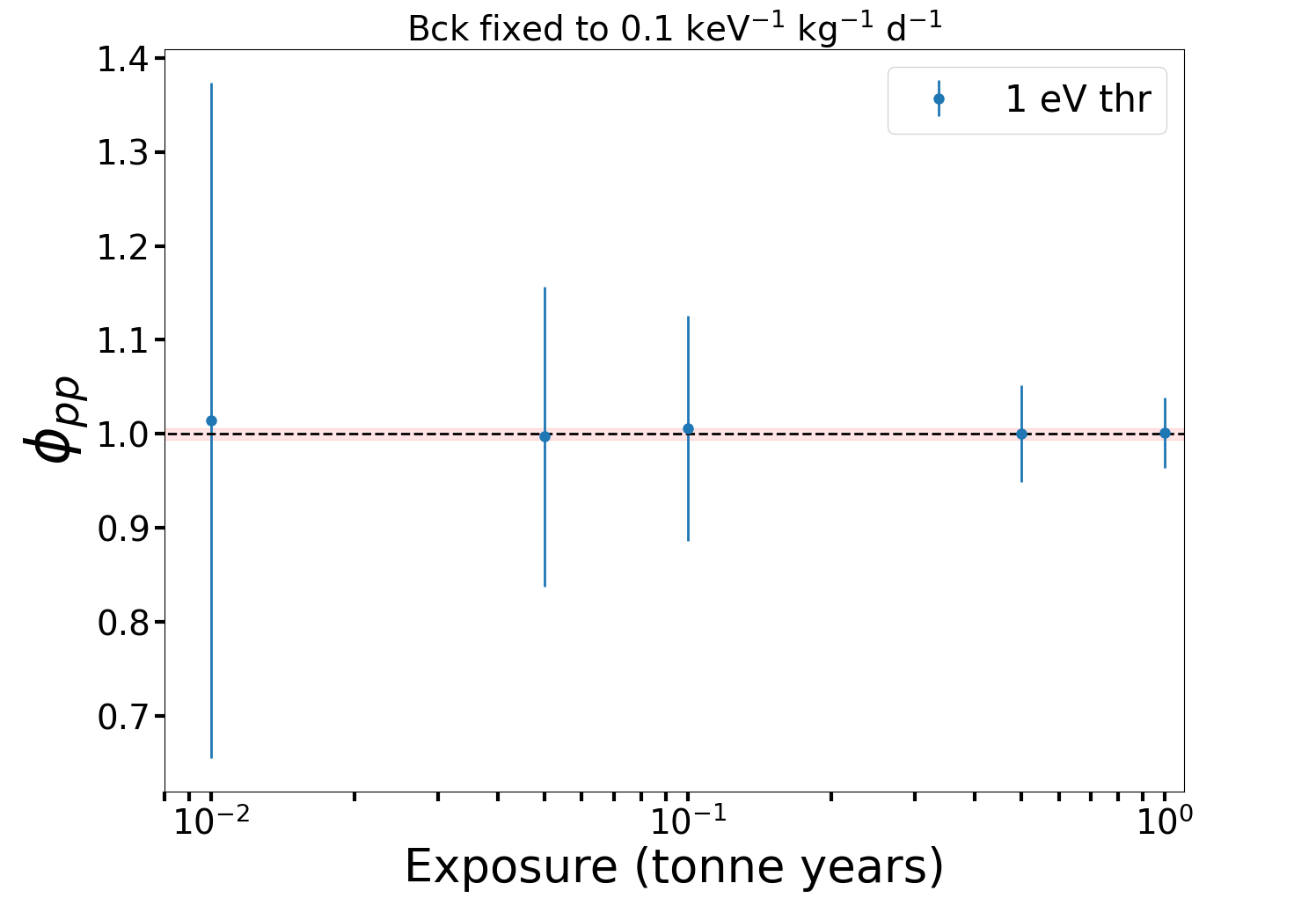}
	\caption{\label{fig:pp_ExpoScan} The data points and error bars represent the mean and standard deviation of the reconstructed pp neutrino flux normalization parameter of 1000 Monte Carlo simulations with different experimental settings. The background rate is fixed at 0.1/(keV$\,$kg$\,$d) and the threshold is fixed at 1$\,$eV for all settings. The exposure is increased from 0.01$\,$tonne$\,$years to 1$\,$tonne$\,$year. The dashed line shows the model input value of the pp flux and the red shaded band represents the theoretical uncertainty.}
\end{figure}

In this case, the flux is correctly reconstructed up to a background level of 1 counts/(keV$\,$kg$\,$d) and starts showing a small bias at the highest background level. The standard deviation of the reconstructed $^{7}$Be flux stays below the theoretical uncertainty of 6$\,\%$ for background levels below about 1$\,$counts/(keV$\,$kg$\,$d) with 6.2$\,\%$ at 1$\,$count/(keV$\,$kg$\,$d). To reach the currently best experimental uncertainty of $^{+2.5}_{-2.7}\,\%$ \cite{Agostini2018}, higher exposures are required, as shown above. 
\\~\\
We test the reconstruction of the pp neutrino flux with a Al$_{2}$O$_{3}$ detector, in which the pp flux is dominant with respect to the $^{7}$Be flux in the energy range between 1$\,$eV and 10$\,$eV. We compare our results with the theoretical uncertainty of 0.6$\,\%$ (Tab. \ref{tab:SolarModel}) and the currently best measurement of the pp flux and its uncertainties 6.1$\,$(1$\,^{+0.096}_{-0.116}$)$\,\cdot\,$10$^{10}\,$cm$^{-2}\,$s$^{-1}$ \cite{Agostini2018}. Due to the low energy depositions of pp neutrinos in the detectors, we consider only a threshold of 1$\,$eV. Again, we fix the background level at a low value of 0.1$\,$counts/(keV$\,$kg$\,$d) and compare the accuracy of the reconstructed values of the pp flux to the theoretical uncertainty as a function of the exposure. The results are shown in Fig. \ref{fig:pp_ExpoScan}.

\begin{figure}[htb!]
	\centering
	\includegraphics[width=0.48\textwidth]{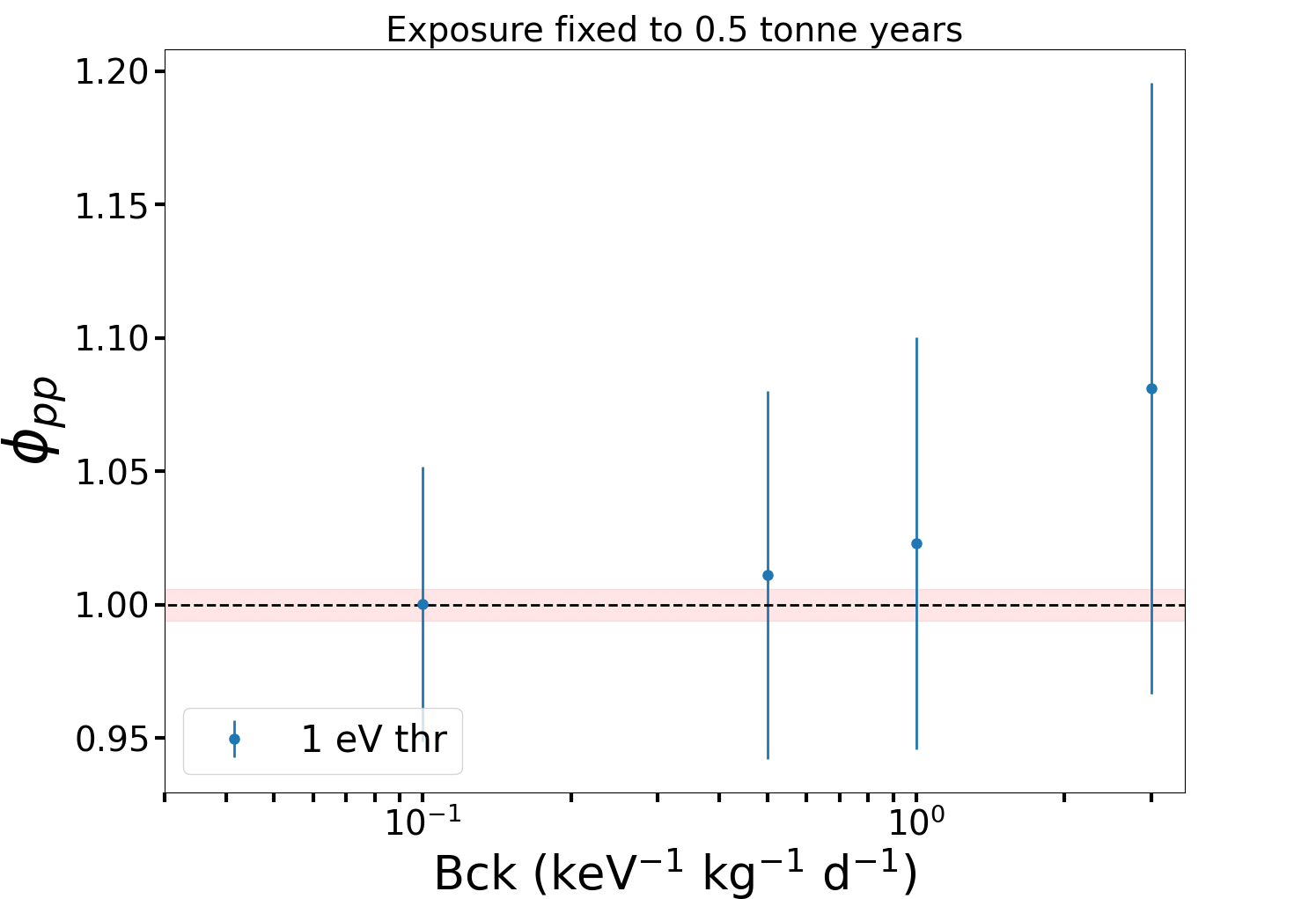}
	\caption{\label{fig:pp_BckScan} The data points and error bars represent the mean and standard deviation of the reconstructed pp neutrino flux normalization parameter of 1000 Monte Carlo simulations. The exposure is fixed at 0.5$\,$tonne$\,$years and the threshold is fixed at 1$\,$eV for all points. The background rate is increased from 0.1/(keV$\,$kg$\,$d) to 3/(keV$\,$kg$\,$d). The dashed line shows the model input value of the pp flux and the red shaded band represents the theoretical uncertainty.}
\end{figure}

The flux is correctly reconstructed for all cases above 0.01$\,$tonne$\,$years. We also investigate the effect of constraining the fluxes of CNO and \textit{hep} neutrinos to the LZ model values for an exposure of 1$\,$tonne$\,$year, in which case the mean of the reconstructed pp flux shows a very small bias of 0.27$\,\%$ above the injected value. The standard deviation of the reconstructed mean value is much larger than the theoretical uncertainty even at high exposures. The theoretical uncertainty is based on the luminosity constraint, which assumes that the nuclear fusion reactions among light elements are responsible for the observed luminosity of the Sun. While this is a very reasonable assumption, this level of precision could not yet be reached by any measurement and is thus not yet experimentally verified. At an exposure of 0.5$\,$tonne$\,$years, the uncertainty is at 5.1$\,\%$, which is below the level of uncertainties reached by current experiments at $^{+9.6}_{-11.6}\,\%$ \cite{Agostini2018}. In the next step, we fix the exposure at 0.5$\,$tonne$\,$years and increase the background rate from 0.1/(keV$\,$kg$\,$d) to 3/(keV$\,$kg$\,$d), shown in Fig. \ref{fig:pp_BckScan}.

\begin{figure}[htb!]
	\centering
	\includegraphics[width=0.48\textwidth]{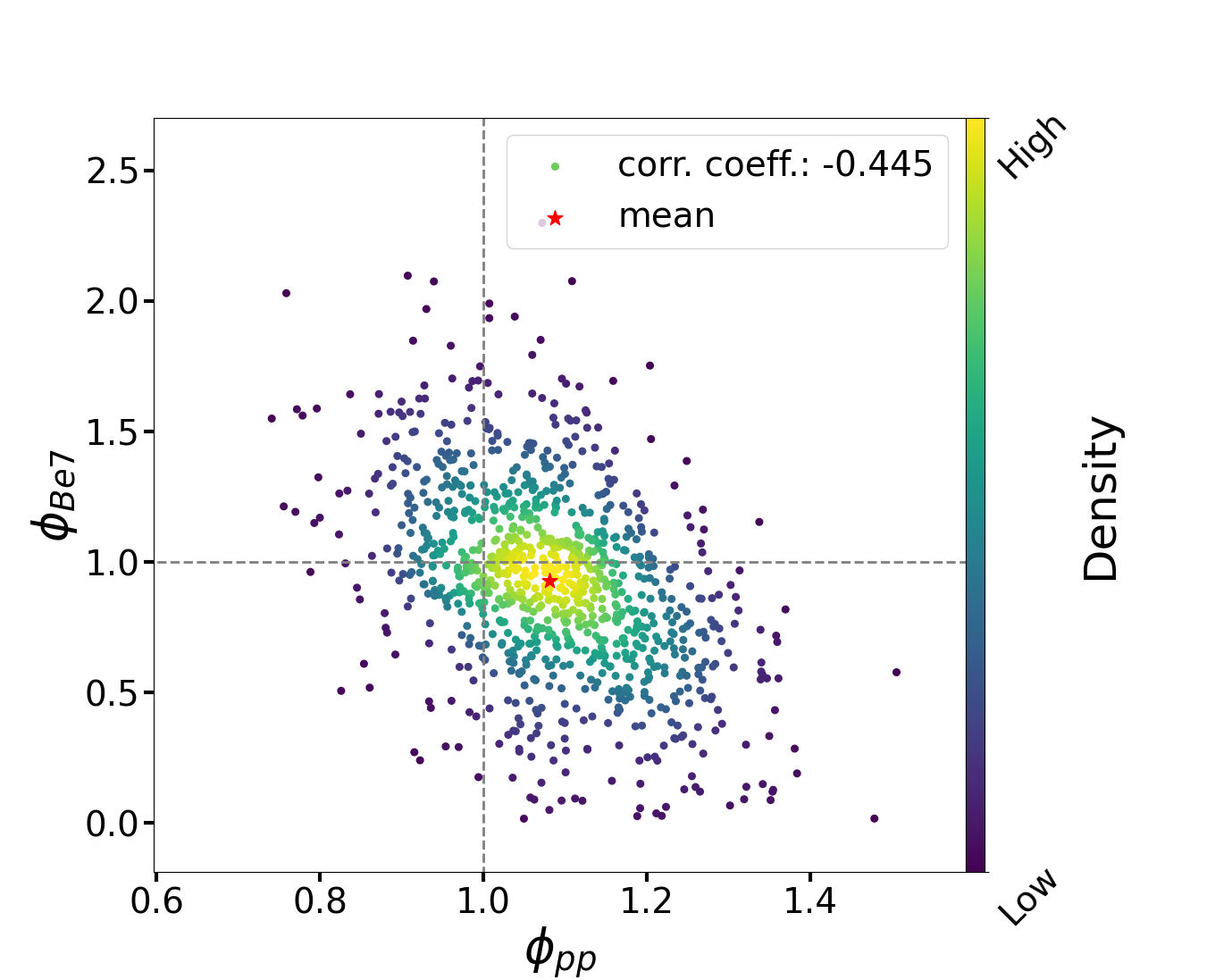}
	\caption{\label{fig:Corr} Anticorrelation between the reconstructed values of the $^{7}$Be and pp fluxes in 1000 Monte Carlo simulations with an exposure of 0.5$\,$tonne$\,$years and a background level of 3/(keV$\,$kg$\,$d) with a pearson correlation coefficient of -0.445. The mean (red star) is shifted towards smaller values for the $^{7}$Be flux and towards larger values for the pp flux.}
\end{figure}

The flux is only reconstructed correctly at the injected value at the lowest background level of 0.1/(keV$\,$kg$\,$d). Above this level, a bias becomes apparent, which increases with an increasing background level. The uncertainty stays below the currently best experimental uncertainty of $^{+9.6}_{-11.6}\,\%$ \cite{Agostini2018} for background levels of up to 1/(keV$\,$kg$\,$d) where the uncertainty is at 7.7$\,\%$. With an increasing background, the anticorrelation between the $^{7}$Be and pp fluxes increases, leading to a systematic overestimation of the reconstruction of the pp flux. Fig. \ref{fig:Corr} shows the anticorrelation between $^{7}$Be and pp for 1000 Monte Carlo simulations with an exposure of 0.5$\,$tonne$\,$years and a background level of 3/(keV$\,$kg$\,$d). The mean of the $^{7}$Be flux is slightly shifted to smaller values, while the mean of the pp flux is shifted towards larger values, showing the bias in the reconstruction of the fluxes at high background levels. An accurate reconstruction of the pp flux parameter is experimentally challenging, requiring very low energy thresholds and background levels as well as large exposures of $ \geq \, \mathcal{O}$(1$\,$tonne$\,$year). \\

Current experiments are still several orders of magnitude away from reaching either the necessary exposure or energy threshold to measure a signal from pp or $^{7}$Be neutrinos via CE$\nu$NS. With a setup of detectors with a threshold of $\mathcal{O}$(eV) and an exposure of $\mathcal{O}$(tonne$\,$years) measurements of these fluxes will become accessible. The measurement of the full flavor independent flux of pp and $^{7}$Be neutrinos are a great test of the current solar models and in combination with results from e.g. Borexino allow for tests of the MSW effect.

\subsection{CNO neutrinos}

For a long time, only upper limits on the CNO neutrino flux could be estimated \cite{Agostini2018}. The first observation of neutrinos produced in the CNO cycle was published in 2020 by the Borexino collaboration \cite{Agostini2020}, which was compatible with both, low-metallicity (LZ) and high-metallicity (HZ) solar models. Two years later, Borexino published an experimentally measured summed flux of the CNO neutrinos of 6.6$\,$(1$\,^{+0.303}_{-0.136}$)$ \, \cdot \,$10$^{8} \,$cm$^{-2} \,$s$^{-1}$ \cite{PhysRevLett.129.252701}. Their results are in agreement with HZ standard solar models, while showing a small tension with LZ models, although only with a moderate statistical significance of about 2$\sigma$. An analysis of the CNO neutrino flux combined with the $^{7}$Be and $^{8}$B solar neutrino fluxes in \cite{PhysRevLett.129.252701} allows to disfavor the B16-AGSS09met (LZ solar model) with respect to the alternative model B16-GS98 (HZ solar model) at 3.1$\sigma$ C.L.
Dark matter experiments reaching the neutrino floor also have the potential to contribute to the resolution of this problem by independently measuring the solar neutrino flux via CE$\nu$NS interactions. We show that the key ingredients needed for such an experiment are low energy thresholds $\mathcal{O}$(eV), which are already reached by current cryogenic solid state experiments, and high exposures $\mathcal{O}$(tonne$\,$year) as well as low background rates $\mathcal{O}$(0.1$\,$keV$^{-1}$kg$^{-1}$d$^{-1}$).

In this section, we restrict the flux parameters of all neutrinos of the pp-chain, with exception of the $^{7}$Be flux, by pull-terms in the likelihood (around their model values and within the theoretical uncertainty of the HZ and LZ models of Tab. \ref{tab:SolarModel}), while the fluxes of the neutrinos of the CNO-cycle are left as fully free parameters. The sensitivity to the individual fluxes of the CNO neutrinos is limited, therefore we combine the sum of the three fluxes into a single fit parameter, $\phi_{\mathrm{CNO}}$, following the approach of Borexino in \cite{PhysRevLett.129.252701}. We test the statistical significance of the fitted summed CNO and $^{7}$Be neutrino fluxes against the hypothesis of a low metallicity solar model (null hypothesis = LZ solar model).

\begin{figure}[htb!]
	\centering
	\includegraphics[width=0.5\textwidth]{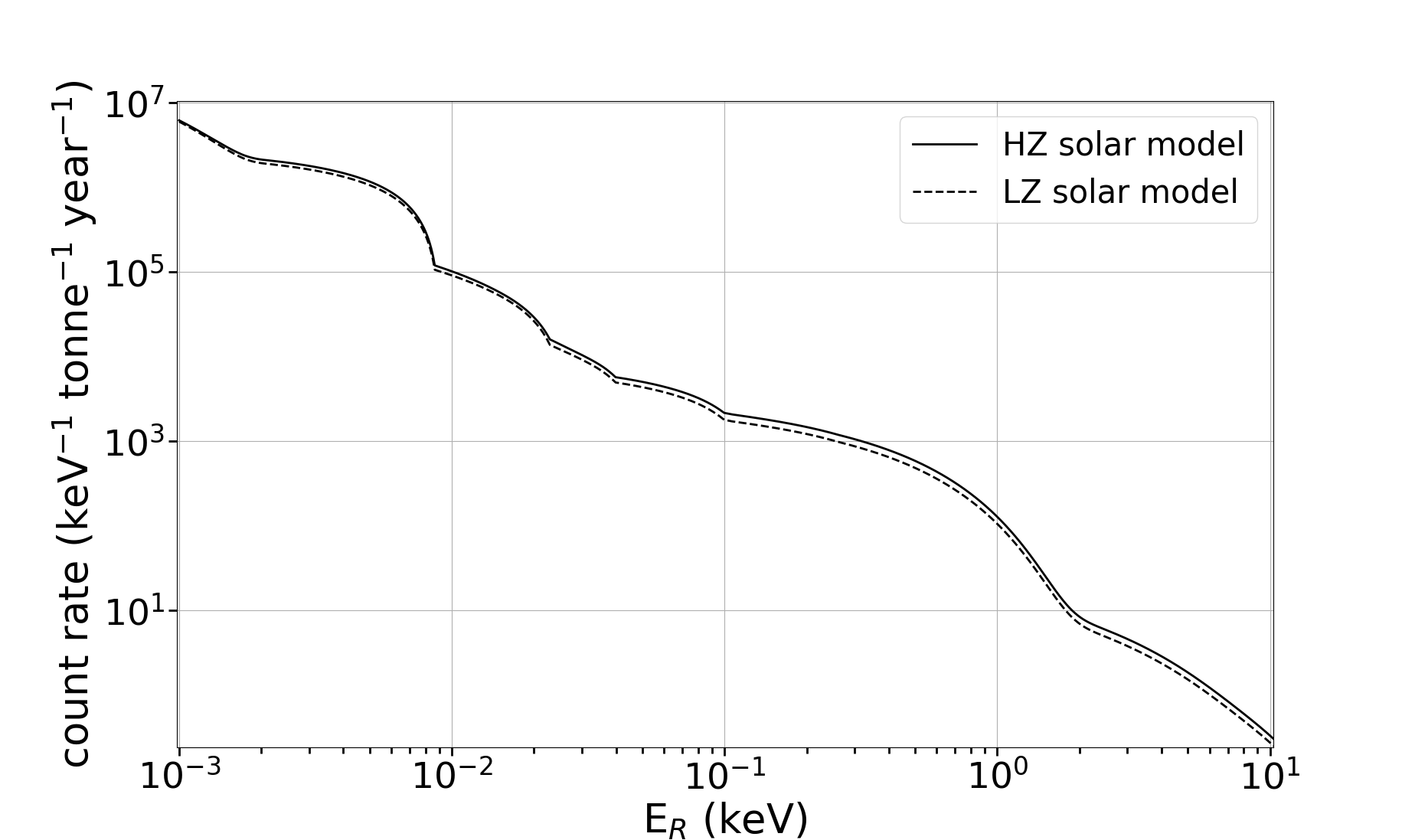}
	\caption{\label{fig:altCNOmodels}Differences in the theoretical differential spectrum of the total solar neutrino flux in the two cases of a HZ solar model and a LZ solar model in a CaWO$_{4}$ detector.}
\end{figure}

The differences between the HZ model and the LZ model (Tab. \ref{tab:SolarModel}) are small. The theoretical differential spectrum of the solar neutrino flux in a CaWO$_{4}$ detector is shown in Fig. \ref{fig:altCNOmodels} for both cases.

As can be seen, the difference between the HZ and LZ models for most neutrino fluxes is 
small ($\sim \,$1$\,\%$) while the flux of $^{7}$Be neutrinos is reduced by a factor $\sim \,$9$\%$, the one of $^{8}$B neutrinos by $\sim \,$18$\, \%$ and the fluxes of $^{13}$N, $^{17}$F and $^{15}$O neutrinos are reduced by roughly 30$\,\%$. The factor of the summed CNO flux parameter and the $^{7}$Be flux is given by:

\begin{equation} \label{eq:CNOfactor}
\begin{split}
	\phi_{\mathrm{CNO,LZ}} & = 0.7194 \cdot \phi_{\mathrm{CNO,HZ}} \\
    \phi_{\mathrm{Be7,LZ}} & = 0.9128 \cdot \phi_{\mathrm{Be7,HZ}}
 \end{split}
\end{equation}

\begin{figure}[htb!]
	\centering
	\includegraphics[width=0.48\textwidth]{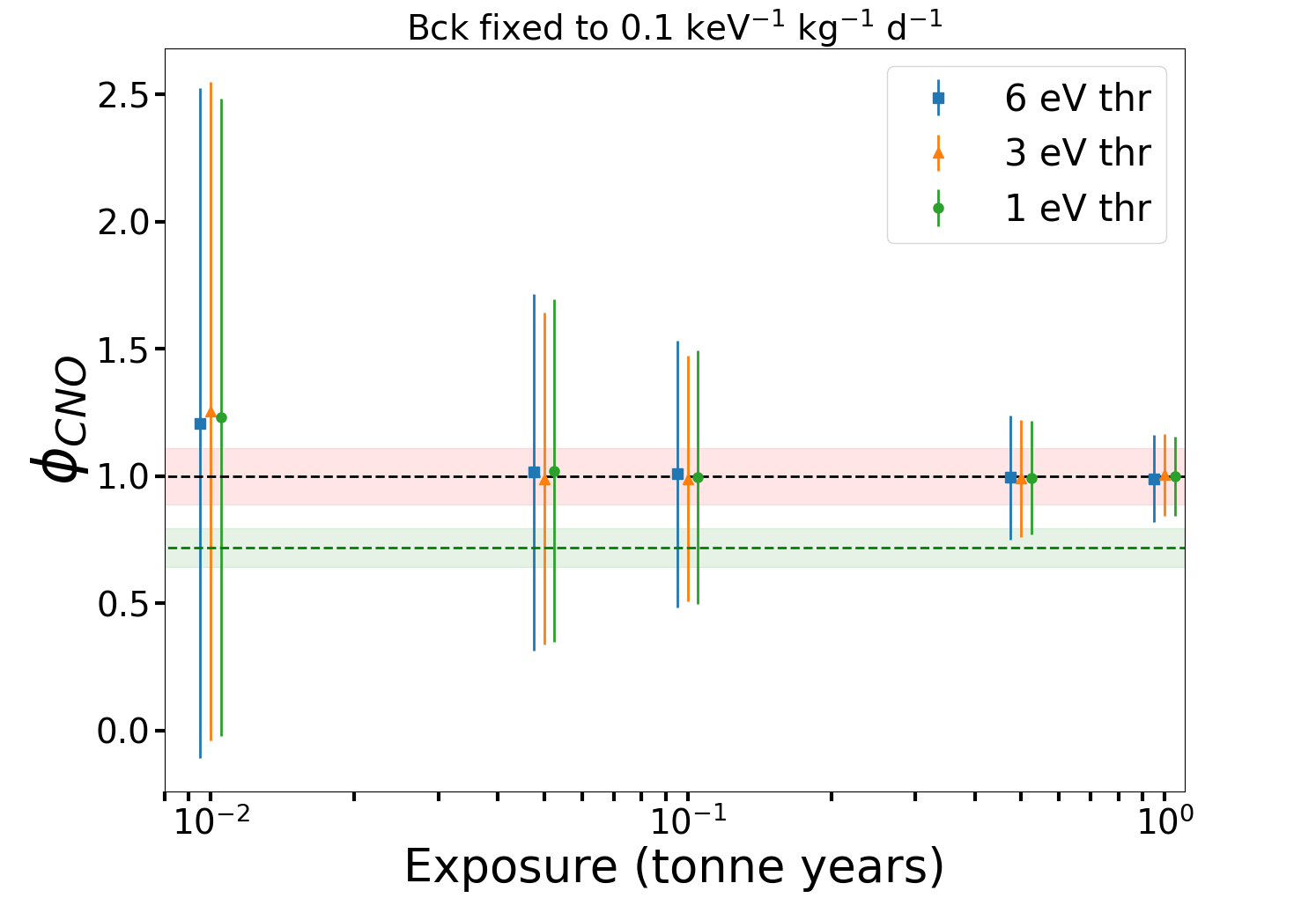}
	\caption{\label{fig:CNO_ExpoScan} The data points and error bars represent the mean and standard deviation of the reconstructed CNO neutrino flux normalization parameter of 1000 Monte Carlo simulations. The background rate is fixed at 0.1/(keV$\,$$\,$d) for all settings. The exposure is increased from 0.01$\,$tonne$\,$years to 1$\,$tonne$\,$year for three different thresholds, indicated by the legend. The data points are slightly shifted around the exposure of the simulations for visibility. The black dashed line and red shaded area show the model input value and uncertainty of the CNO flux of the HZ solar model and the green dashed line and green shaded area show the model value and uncertainty of the CNO flux of the LZ solar model.}
\end{figure}

\begin{figure}[htb!]
	\centering
	\includegraphics[width=0.48\textwidth]{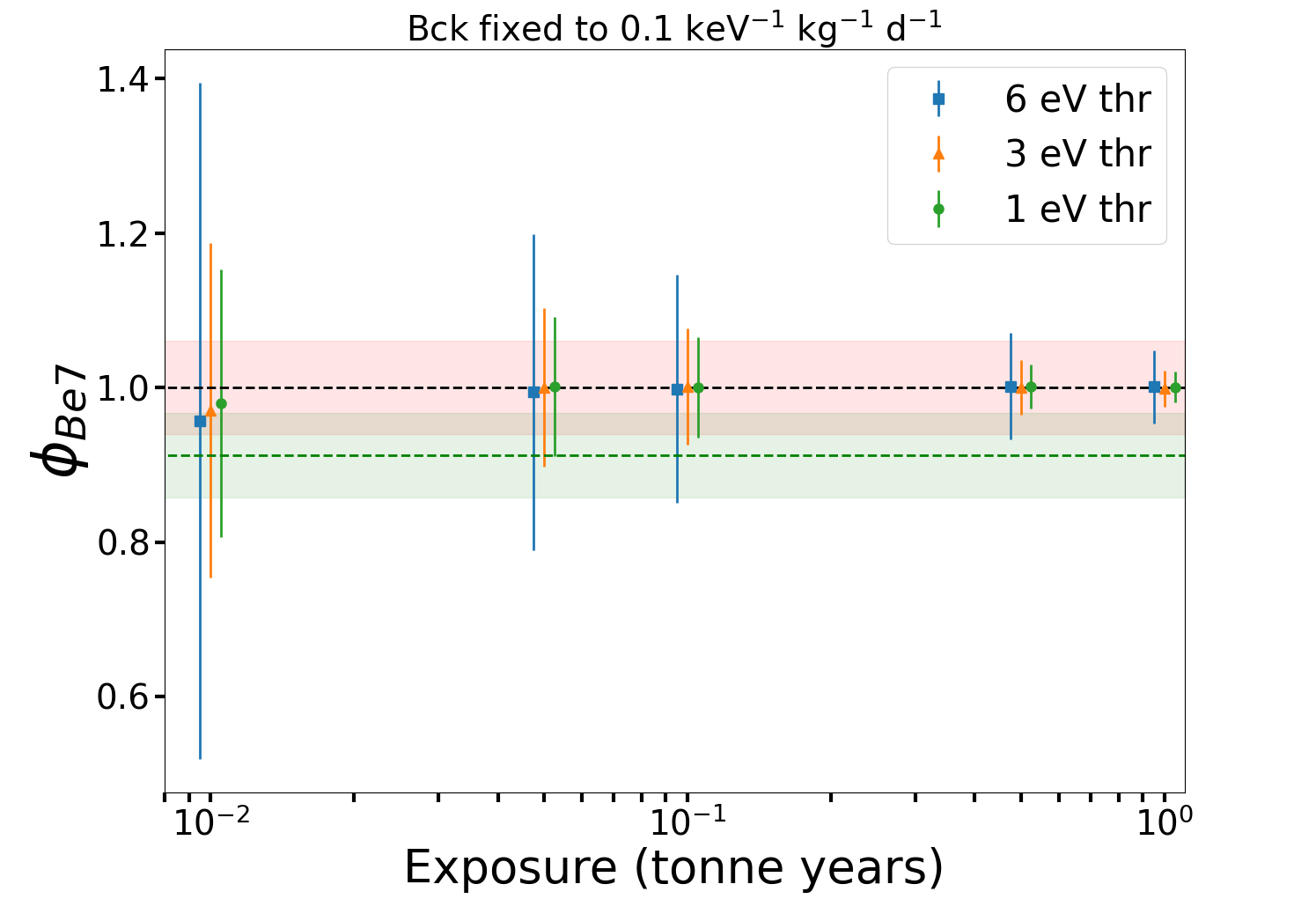}
	\caption{\label{fig:Be7met_ExpoScan} The data points and error bars represent the mean and standard deviation of the reconstructed $^{7}$Be neutrino flux normalization parameter of 1000 Monte Carlo simulations. The background rate is fixed at 0.1/(keV$\,$$\,$d) for all settings. The exposure is increased from 0.01$\,$tonne$\,$years to 1$\,$tonne$\,$year for three different thresholds, indicated by the legend. The data points are slightly shifted around the exposure of the simulations for visibility. The black dashed line and red shaded area show the model input value and uncertainty of the $^{7}$Be flux of the HZ solar model and the green dashed line and green shaded area show the model value and uncertainty of the $^{7}$Be flux of the LZ solar model.}
\end{figure}

Analogous to the previous section, we fix the background level at 0.1$\,$counts/(keV$\,$kg$\,$d). We compare the mean value and the standard deviation of the reconstructed CNO flux and $^{7}$Be flux of 1000 Monte Carlo simulations of the HZ solar model in a CaWO$_{4}$ detector to the expected values of the HZ and the LZ solar model as a function of the exposure for three different thresholds (1$\,$eV, 3$\,$eV, 6$\,$eV). The theoretical uncertainty on the full CNO flux is 11.13$\,\%$ in case of the HZ model and 10.45$\,\%$ in case of the LZ model. The results for the CNO flux are shown in Fig. \ref{fig:CNO_ExpoScan} and for the $^{7}$Be flux in Fig. \ref{fig:Be7met_ExpoScan}.

\begin{figure}[htb!]
	\centering
	\includegraphics[width=0.48\textwidth]{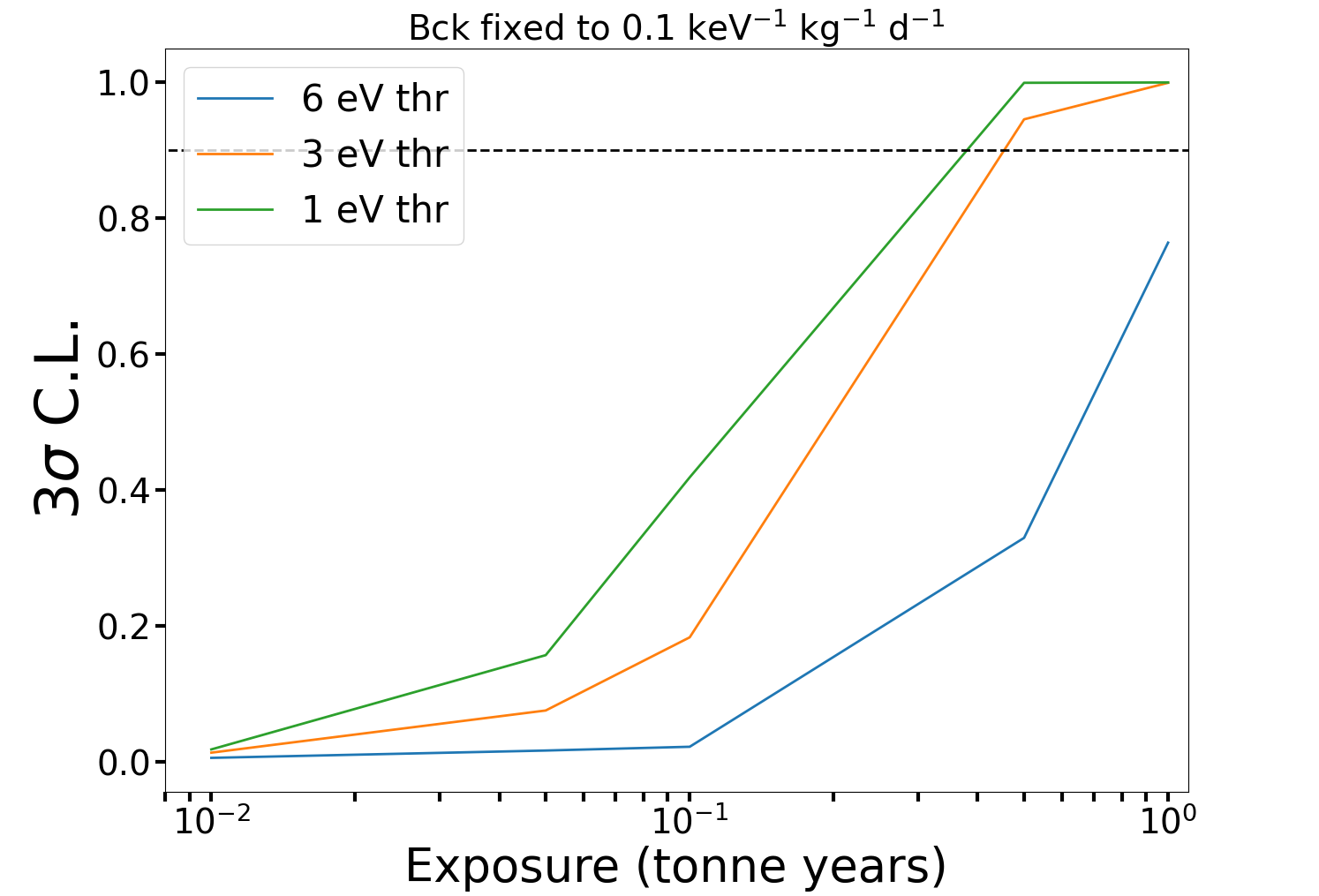}
	\caption{\label{fig:CNO_ExpoScan_Sig} The lines show the fraction of simulated datasets that exclude the null hypothesis of the LZ solar model with a significance of at least 3$\sigma$ as a function of the exposure. The background rate is fixed at 0.1/(keV$\,$kg$\,$d) for all settings. The exposure is increased from 0.01$\,$tonne$\,$years to 1$\,$tonne$\,$year for three different thresholds, indicated by the legend. The dashed line shows the 90$\,$\% probability of reaching a C.L. of 3$\sigma$.}
\end{figure}

The flux of CNO and $^{7}$Be neutrinos is correctly reconstructed in all cases above an exposure of 0.01$\,$tonne$\,$years. Up to an exposure of 0.1$\,$tonne$\,$years, the standard deviation of the reconstructed values overlap with the expected value of a LZ solar model for all three thresholds in the case of CNO neutrinos and for a 6$\,$eV threshold in the case of $^{7}$Be neutrinos. There is no strong dependence on the threshold for the reconstruction of CNO neutrinos, since they become relevant above an energy of about 10$\,$eV (see Fig. \ref{fig:SolNuRecoilSpec}). With increasing exposure, the uncertainties get smaller for all three thresholds and do not overlap. At an exposure of 0.5$\,$tonne$\,$years the uncertainties on the reconstructed CNO flux are at 22.3$\,\%$, 23.0$\,\%$, 24.3$\,\%$ for the thresholds 1$\,$eV, 3$\,$eV, 6$\,$eV, respectively. This is comparable to the currently best experimental uncertainty of $^{+30.3}_{-13.6}\,\%$ \cite{PhysRevLett.129.252701}, while still above the theoretical uncertainty on the HZ CNO flux of 11.13$\,\%$. The uncertainties on the reconstructed value of the $^{7}$Be flux with an exposure of 0.5$\,$tonne$\,$years  are 2.9$\,\%$, 3.5$\,\%$, 6.9$\,\%$ for the thresholds 1$\,$eV, 3$\,$eV, 6$\,$eV, respectively (theoretical uncertainty: 6$\,\%$, currently best experimental uncertainty: $^{+2.5}_{-2.7}\,\%$ \cite{Agostini2018}). To express this in terms of a statistical significance we perform a hypothesis test. The null hypothesis describes the LZ solar model, with $\phi_{\mathrm{CNO}} = 0.7194$ and $\phi_{Be7} = 0.9128$, leading to the following test statistic, $q_{\mathrm{LZ}}$:


\begin{equation} \label{eq:LZCNOTStat}
	q_{\mathrm{LZ}} = -2 \ln\left(\frac{\mathcal{L}_{\mathrm{LZ}}(\phi_{\mathrm{CNO}},\,\phi_{Be7},\,\hat{\hat{\vec{b}}})}{\mathcal{L}_{\mathrm{HZ}}(\hat{\phi}_{\mathrm{CNO}},\,\hat{\phi}_{\mathrm{Be7}},\,\hat{\vec{b}})}\right)
\end{equation}

with the background parameters, $\vec{b}$, being all remaining neutrino fluxes of the pp-chain, $\vec{\phi}_{pp-\nu}$, and the rate of the flat background, $r_{\mathrm{b}}$. In the term in the numerator ($\mathcal{L}_{\mathrm{LZ}}$), $\phi_{\mathrm{CNO}}$ and $\phi_{Be7}$ are fixed to the respective LZ values, while the remaining neutrino fluxes are constrained by the LZ model. In the likelihood in the denominator ($\mathcal{L}_{\mathrm{HZ}}$), $\hat{\phi}_{\mathrm{CNO}}$ and $\hat{\phi}_{Be7}$ are free fit parameters, while the remaining neutrino fluxes are constrained by the HZ model. We perform 10$^{4}$ Monte Carlo simulations for each experimental setting under the LZ model to create the distribution of the test statistic under the null hypothesis and empirically determine the 99.73 percentile, $q_{3\sigma}$, corresponding to the value at which we reject the LZ model a statistical significance of 3$\sigma$. 
We generate 4000 Monte Carlo datasets for each experimental setting under the HZ solar model and calculate the corresponding observed value of the test statistic of eq. \ref{eq:LZCNOTStat}. Figure \ref{fig:CNO_ExpoScan_Sig} shows the fraction of simulated datasets that exclude the null hypothesis of the LZ solar model with a significance of at least 3$\sigma$ ($q_{\mathrm{LZ}} \geq q_{3\sigma}$) as a function of the increasing exposure.

\begin{figure}[htb!]
	\centering
	\includegraphics[width=0.48\textwidth]{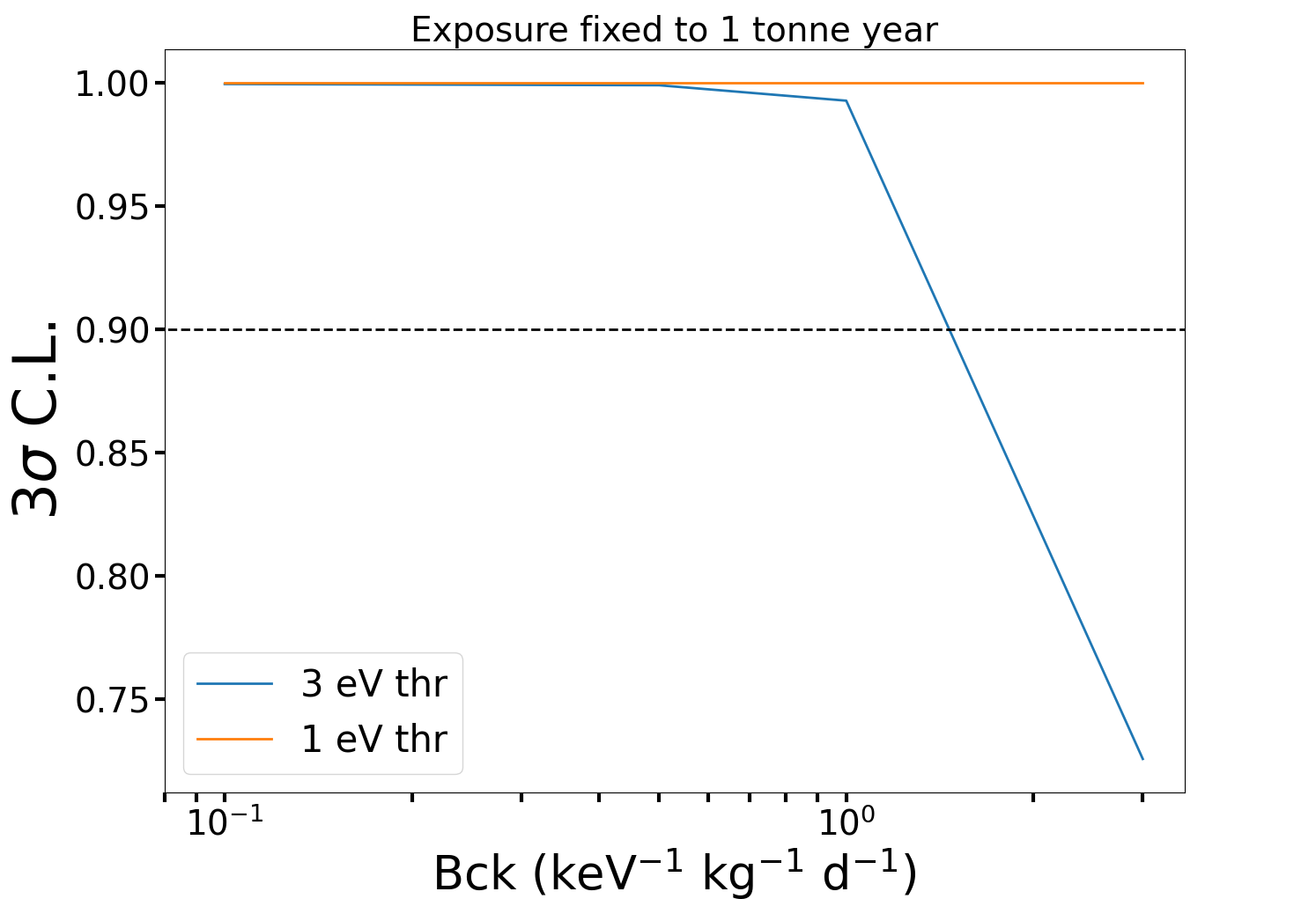}
	\caption{\label{fig:CNO_BckScan_Sig} The lines show the fraction of simulated datasets that exclude the null hypothesis of the LZ solar model with a significance of at least 3$\sigma$ as a function of the background rate. The exposure is fixed at 1 tonne year for all settings. The background rate is increased from 0.1/(keV$\,$kg$\,$d) to 3/(keV$\,$kg$\,$d) for two different thresholds, indicated by the legend. The dashed line shows the 90$\,$\% probability of reaching a C.L. of 3$\sigma$.}
\end{figure}

The LZ solar model can be excluded with a significance of 3$\sigma$ at a confidence level of at least 90$\,$\% for the settings with an exposure of 0.5$\,$tonne$\,$years or larger in the case of a threshold of 3$\,$eV or lower. With a threshold above 3$\,$eV, the benchmark of a 90$\,\%$ probability is not reached even at the highest tested exposure of 1$\,$tonne$\,$year. We therefore drop the setting with a threshold of 6$\,$eV in the next step, in which we fix the exposure at 1$\,$tonne$\,$year, while the background rate is increased from 0.1/(keV$\,$kg$\,$d) to 3/(keV$\,$kg$\,$d) for the thresholds of 1$\,$eV and 3$\,$eV. Analogous to before, we plot the C.L. at which the LZ solar model can be rejected with a significance of 3$\sigma$ against the increasing background rate in Fig. \ref{fig:CNO_BckScan_Sig}.

\begin{figure}[htb!]
	\centering
	\includegraphics[width=0.48\textwidth]{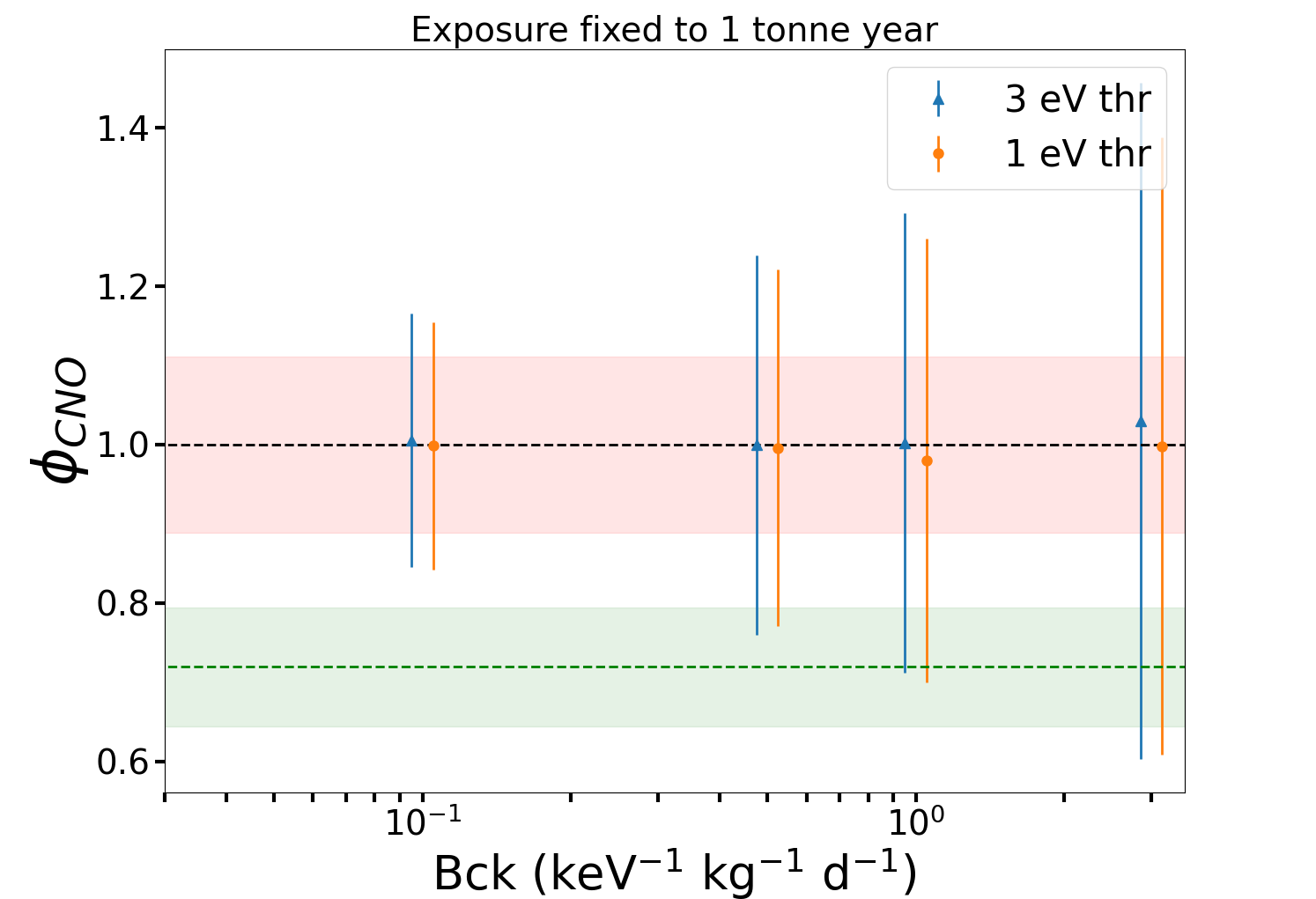}
	\caption{\label{fig:CNO_BckScan} The data points and error bars represent the mean and standard deviation of the reconstructed CNO neutrino flux normalization parameter of 1000 Monte Carlo simulations. The exposure is fixed at 1$\,$tonne$\,$year for all settings. The background rate is increased from 0.1/(keV$\,$kg$\,$d) to 3/(keV$\,$kg$\,$d) for two different thresholds, indicated by the legend. The data points are slightly shifted around the background rate of the simulations for visibility. The black dashed line and red shaded area show the model input value and uncertainty of the CNO flux of the HZ solar model and the green dashed line and green shaded area show the model value and uncertainty of the CNO flux of the LZ solar model.}
\end{figure}

\begin{figure}[htb!]
	\centering
	\includegraphics[width=0.48\textwidth]{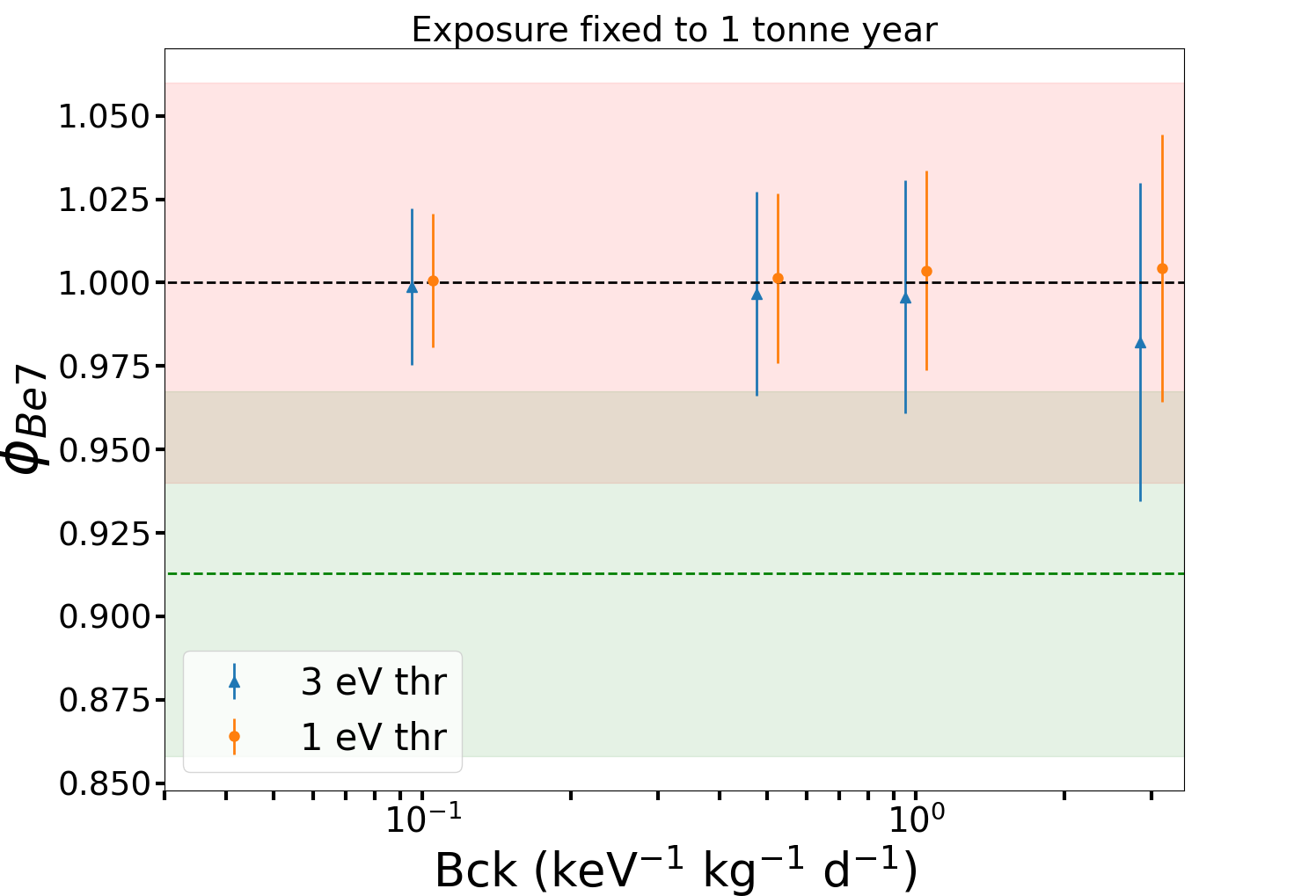}
	\caption{\label{fig:Be7met_BckScan} The data points and error bars represent the mean and standard deviation of the reconstructed $^{7}$Be neutrino flux normalization parameter of 1000 Monte Carlo simulations. The exposure is fixed at 1$\,$tonne$\,$year for all settings. The background rate is increased from 0.1/(keV$\,$kg$\,$d) to 3/(keV$\,$kg$\,$d) for two different thresholds, indicated by the legend. The data points are slightly shifted around the background rate of the simulations for visibility. The black dashed line and red shaded area show the model input value and uncertainty of the $^{7}$Be flux of the HZ solar model and the green dashed line and green shaded area show the model value and uncertainty of the $^{7}$Be flux of the LZ solar model.}
\end{figure}

In the case of the lowest threshold of 1$\,$eV, the probability of rejecting the LZ model with a significance of at least 3$\sigma$ stays well above the 90$\,\%$ benchmark for all tested background levels. In the case of the 3$\,$eV threshold, the probability at which the LZ solar model can be rejected at 3$\sigma$ significance drops below 90$\,$\% at a background level between 1/(keV$\,$kg$\,$d) and 3/(keV$\,$kg$\,$d). Analogous to Fig. \ref{fig:CNO_ExpoScan} and Fig. \ref{fig:Be7met_ExpoScan}, we also plot the reconstructed values of the CNO flux and their standard deviation against the increasing background level in Fig. \ref{fig:CNO_BckScan} and the ones of the $^{7}$Be flux in Fig. \ref{fig:Be7met_BckScan}.

The mean of the reconstructed values of the CNO flux stays close to the model input value at all background levels. In the cases of a background rate smaller than 0.5/(keV$\,$kg$\,$d), the standard deviation of the reconstructed values of the CNO flux does not overlap with the expected value of the LZ model. Furthermore, with values of 22.5$\,\%$ and 24.0$\,\%$ for the thresholds of 1$\,$eV and 3$\,$eV, the uncertainties stay comparable to the experimental ones of $^{+30.3}_{-13.6}\,\%$ \cite{PhysRevLett.129.252701}, but above the theoretical one of 11.13$\,\%$ for the HZ CNO flux. Above this background level, the reconstructed values of the CNO flux start to strongly overlap with the LZ model. The reconstructed values of the $^{7}$Be flux are well separated from the LZ model for background levels of up to 1/(keV$\,$kg$\,$d) with uncertainties of 3.0$\,\%$ and 3.5$\,\%$ for the thresholds of 1$\,$eV and 3$\,$eV, staying below the theoretical uncertainty of 6$\,\%$ and slightly above the experimental one of $^{+2.5}_{-2.7}\,\%$ \cite{Agostini2018}. In the case of a low threshold of 1$\,$eV, the reconstructed values don't overlap with the LZ model up to the largest tested background rate of 3/(keV$\,$kg$\,$d) with an uncertainty of 4.0$\,\%$, while the values obtained with a threshold of 3$\,$eV start start showing a bias, which is leaking into the LZ model region, reflecting the drop of the probability in Fig. \ref{fig:CNO_BckScan_Sig}.

These results show that the capability of rejecting the LZ solar model is mainly driven by the ability to reconstruct the $^{7}$Be flux. For thresholds larger than about 6$\,$eV, there is not enough sensitivity to reject the LZ solar model. If an experiment with an exposure $\mathcal{O}$(1$\,$tonne$\,$year) achieves a threshold of about 3$\,$eV or below and the background rate stays below levels of $\mathcal{O}$(1/(keV$\,$kg$\,$d)), it has a high probability ($\geq \,$90$\,$\%) to reject the LZ solar model with at least 3$\sigma$ significance.

\section{Sensitivity to dark matter}\label{DM_Sens}
In the second part of this work we want to investigate the influence of a solar neutrino background on the sensitivity to dark matter signals. We first give a short description of the definition of the neutrino floor, followed by the calculation of sensitivity limits for dark matter signals below the neutrino floor.

\subsection{The neutrino floor}\label{Nufog}

There is no unique definition of the neutrino floor. A common choice is described in \cite{PhysRevD.89.023524}: For a large number of different energy thresholds, a background-free exclusion limit (defined as isovalues corresponding to zero observed WIMP events) is calculated at 90$\,$\% C.L. The exposure at which each of these exclusion lines is calculated is adjusted so that exactly one neutrino background event is expected. By defining the neutrino floor as the minimum of these exclusion limits at each dark matter particle mass, the best possible estimate for the dark matter sensitivity for each dark matter mass in the presence of a single neutrino event can be calculated. 

\begin{figure}[h]
	\centering
	\includegraphics[width=0.48\textwidth]{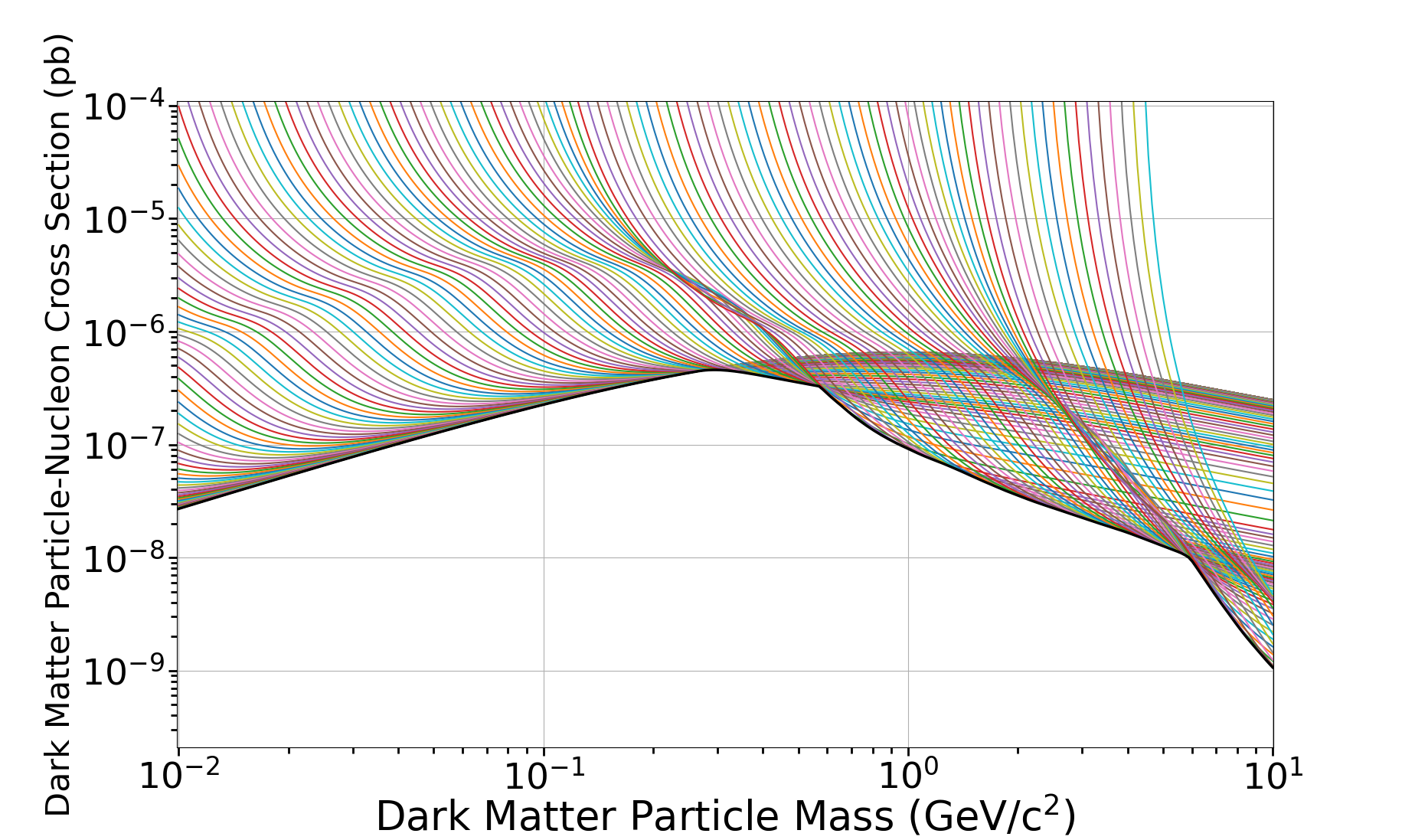}
	\caption{\label{fig:NufloorEx}Set of background-free 90$\,$\% C.L. exclusion limits for exposures corresponding to one neutrino event above the energy threshold for 200 different thresholds between 0.1$\, \mu$eV and 10$\,$keV in CaWO$_{4}$. The black line is constructed by taking the minimum of the individual exclusion limits at each mass.}
\end{figure}

We show an example of the neutrino floor calculation for a CaWO$_{4}$ detector in Fig. \ref{fig:NufloorEx} for a dark matter mass range from 10$\,$MeV/c$^{2}$ to 10$\,$GeV/c$^{2}$. The calculation was done for 200 different thresholds ranging from 0.1$\, \mu$eV to 10$\,$keV. The minimum of all lines at each mass, shown as a black line, represents the neutrino floor according to the definition we adopted.

\subsection{Sensitivity limits}

We calculate sensitivity limits for a dark matter discovery in the presence of a solar neutrino background for different experimental settings, defined by the target material, energy threshold and exposure. The limits we calculate in this section correspond to a 90$\,$\% C.L. limit for a 3$\sigma$ dark matter discovery. This means that a dark matter particle with mass and scattering cross section above the calculated limit would lead to a discovery with at least 3$\sigma$ significance in at least 90$\,$\% of the experiments with the given settings. Some important points have to be mentioned here: We consider no other backgrounds than neutrinos and no energy resolution in the simulations. Furthermore, we assume an (unrealistic) energy independent efficiency of a 100$\,$\% for the experiments. Hence, the limits represent a best case scenario. The experimental settings for which the limits are computed are given in Tab. \ref{tab:SimSettings}.

\begin{table}[h]
\centering
	\caption{\label{tab:SimSettings}A 3$\sigma$ discovery limit is calculated for each of these experimental settings.}
	\newcolumntype{C}{>{\centering\arraybackslash}X}
	\setlength\extrarowheight{3pt}
	\noindent
    \begin{tabularx}{0.48\textwidth}{ C C C}
    \hline
    Material & Threshold (eV) & Exposure (tonne $\cdot$ year) \\ \hline
    CaWO$_{4}$ & 1 & 0.1 \\
    CaWO$_{4}$ & 1 & 1 \\
    CaWO$_{4}$ & 0.1 & 1 \\
    CaWO$_{4}$ & 1 & 10 \\
    Al$_{2}$O$_{3}$ & 1 & 0.1 \\
    Al$_{2}$O$_{3}$ & 1 & 1 \\
    Al$_{2}$O$_{3}$ & 1 & 10 \\ \hline
    \end{tabularx}
\end{table}

We assume an energy threshold of 1$\,$eV in all experiments. For the setting with an exposure of 1$\,$tonne$\,$year with CaWO$_{4}$ detectors we also calculate a limit for a threshold of only 0.1$\,$eV to show the strong effect of a reduced threshold in the low mass regime. \\

\paragraph{Method:} We use the same likelihood function as defined in eq. \ref{eq:BinnedL}, with some modifications. We replace the background terms $f_{\mathrm{b}}$ in eq. \ref{eq:PDFNorm} - \ref{eq:UnNormPDF} with a term corresponding to a dark matter signal, $f_{\chi}$, which we define as:

\begin{equation}
    \tilde{f_{\chi}}(E_{\mathrm{R}},\sigma_{\chi}) = \epsilon \cdot \sigma_{\chi} \cdot \dfrac{\mathrm{d}R_{\chi}}{\mathrm{d}E_{\mathrm{R}}}(E_{\mathrm{R}},\sigma_{\chi} = 1 \, \mathrm{pb})
\end{equation}

with $\frac{\mathrm{d}R_{\chi}}{\mathrm{d}E_{\mathrm{R}}}(E_{\mathrm{R}},\sigma_{\chi} = 1 \, \mathrm{pb})$ being the expected differential recoil rate of dark matter with a dark matter-nucleon cross section of 1$\,$pb, making the fit parameter $\sigma_{\chi}$ the cross section in units of pb. The expected differential recoil spectrum of dark matter is briefly sketched out in App. \ref{DM_SolNu_models}. As now we are interested in a fit of the dark matter signal, while considering solar neutrinos as a background, all flux parameters are constrained by pull-terms around their theoretical HZ model values of Tab. \ref{tab:SolarModel}.

We quantify the statistical significance of the fit of a dark matter signal in the presence of a background (alternative hypothesis, $H_{1}$), compared to a background only hypothesis (null hypothesis, $H_{0}$), with a profile likelihood ratio test statistic, defined as $q_{0}$:

\begin{equation} \label{eq:NufloorTStat}
	q_{0} = \begin{cases}
	-2 \cdot \ln\left(\frac{\mathcal{L}(\sigma_{\chi}=0,\hat{\hat{\vec{\phi}}}_{\nu})}{\mathcal{L}(\hat{\sigma}_{\chi},\hat{\vec{\phi}}_{\nu})}\right) &\quad ,\hat{\sigma}_{\chi} \geq 0 \\
	0 &\quad ,\hat{\sigma}_{\chi} < 0 \\
	\end{cases}
\end{equation}

The parameters are the dark matter-nucleon interaction cross section, $\sigma_{\chi}$, and eight constrained normalization parameters, $\vec{\phi}_{\nu}$, describing the neutrino fluxes of the solar neutrinos.

The distribution of $q_{0}$ should follow a half $\chi^{2}$ distribution with one degree of freedom under the null hypothesis ($\sigma_{\chi}=0$). This has been investigated and can be confirmed for all dark matter masses in the mass range investigated in this section. Therefore, the statistical significance in units of sigma can be expressed as the square root of the observed test statistic ($Z = \sqrt{q_{0}^{\mathrm{obs}}}$).

We calculate the limits for a dark matter particle mass range from 50$\,$MeV/c$^{2}$ to 10$\,$GeV/c$^{2}$. For a given mass point, we define an interval of cross sections. For each cross section in this interval, we generate 1000 Monte Carlo datasets and subsequently fit them with both likelihood functions ($\mathcal{L}_{\mathrm{H}_{0}}$ and $\mathcal{L}_{\mathrm{H}_{1}}$), from which we compute the observed test statistic $q_{0}^{\mathrm{obs}}$ (eq. \ref{eq:NufloorTStat}). We interpolate the fraction of $Z = \sqrt{q_{0}^{\mathrm{obs}}} \geq 3$ as a function of the cross sections in the defined interval, from which we can determine the value of the cross section $\sigma_{\chi}$ at which $Z \geq 3$ in 90$\,$\% of the experiments. \\

\begin{figure}[h]
	\centering
	\includegraphics[width=0.5\textwidth]{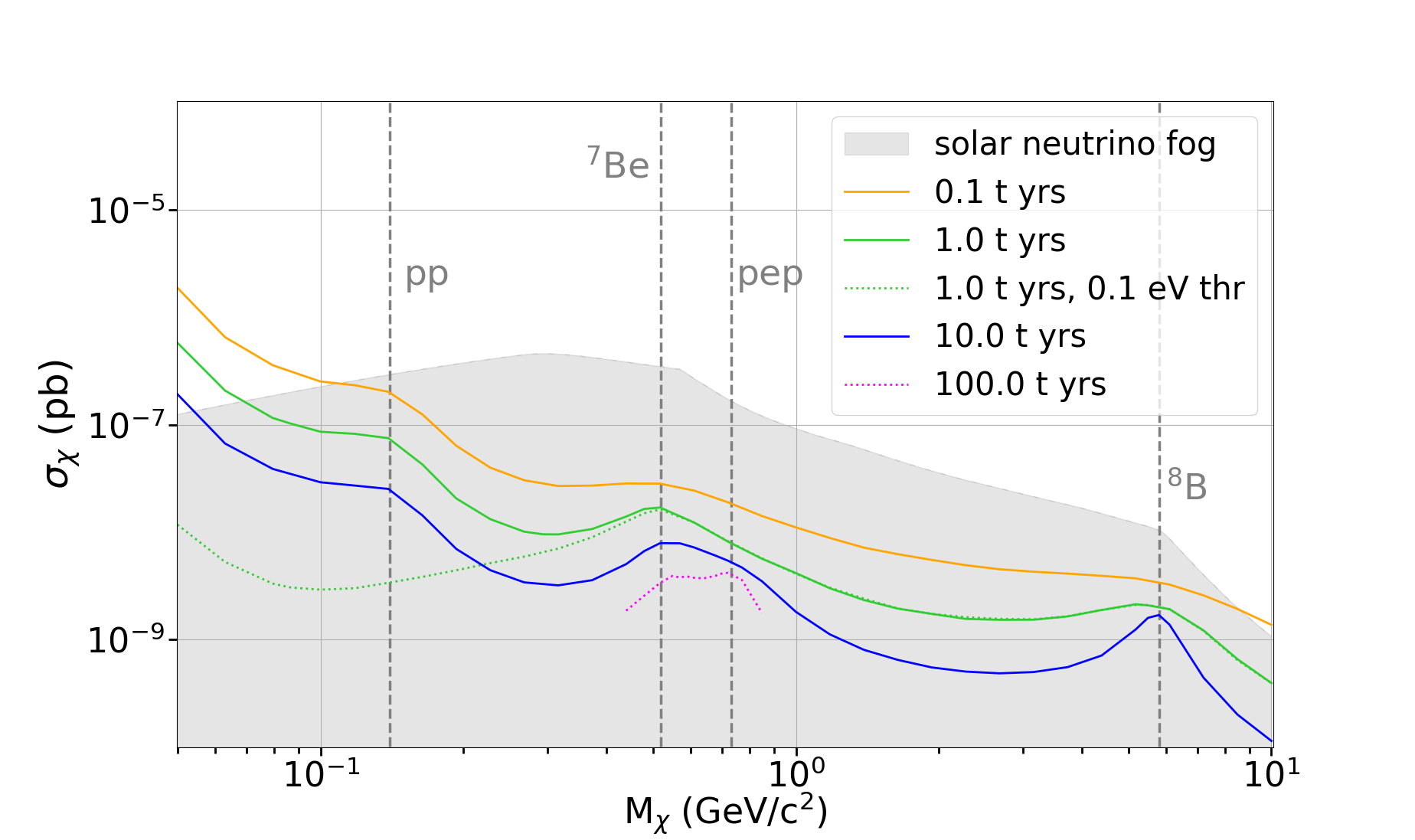}
	\caption{\label{fig:CaWO4_SensLimits}3$\sigma$ discovery potential for dark matter in different experimental settings in CaWO$_{4}$ detectors. The dotted green line is calculated with a threshold of 0.1$\,$eV. All other limits are calculated with a threshold of 1$\,$eV. The dotted magenta line is calculated in a limited dark matter mass range with an exposure of 100$\,$tonne$\,$years. The vertical gray dashed lines mark the dark matter masses at which the sensitivity is limited due to the strong similarities of the expected dark matter spectrum to one of the solar neutrino components. The gray shaded area depicts the neutrino fog.}
\end{figure}

\paragraph{Results:} The resulting dark matter discovery potentials for CaWO$_{4}$ detectors is shown in Fig. \ref{fig:CaWO4_SensLimits}. The sensitivity limits fully immerse into the neutrino floor. This shows that the neutrino floor is not a hard limit and a dark matter discovery is still possible even below the classic definition. For this reason the term neutrino floor is often replaced by neutrino fog.

We additionally calculate the limit for an exposure of 1 tonne year and an even lower threshold of only 0.1$\,$eV (green dotted line in Fig. \ref{fig:CaWO4_SensLimits}). For dark matter masses above about 500$\,$MeV/c$^{2}$, this limit is identical to the one calculated for a threshold of 1$\,$eV with the same exposure. At lower masses, the lower threshold starts to drastically increase the sensitivity towards smaller cross sections. Below a mass of about 140$\,$MeV/c$^{2}$, the limit calculated with a threshold of 0.1$\,$eV reaches cross sections more than one order of magnitude smaller than the limit calculated with a threshold of 1$\,$eV. Moreover, below masses of about 200$\,$MeV/c$^{2}$ the sensitivity limit reaches lower cross sections than the limit calculated with an exposure of 10$\,$tonne$\,$years. This shows that, in order to increase sensitivity to light dark matter masses in the presence of a solar neutrino background, decreasing the energy threshold is much more effective than increasing the exposure.

Generally, below the neutrino floor the sensitivity increases with the square root of the exposure ($\sim \, \sqrt{\epsilon}$), corresponding to a poissonian background substraction regime. This behaviour was already observed e.g. in \cite{PhysRevD.90.083510}. The dashed vertical lines in Fig. \ref{fig:CaWO4_SensLimits} mark the dark matter masses at which the gain in sensitivity with an increased exposure deviates from this behaviour. The reason for this deviation is the strong similarity of the expected dark matter spectra with a particular component of the solar neutrino spectrum (indicated at the dashed lines).

\begin{figure}[h]
	\centering
	\includegraphics[width=0.48\textwidth]{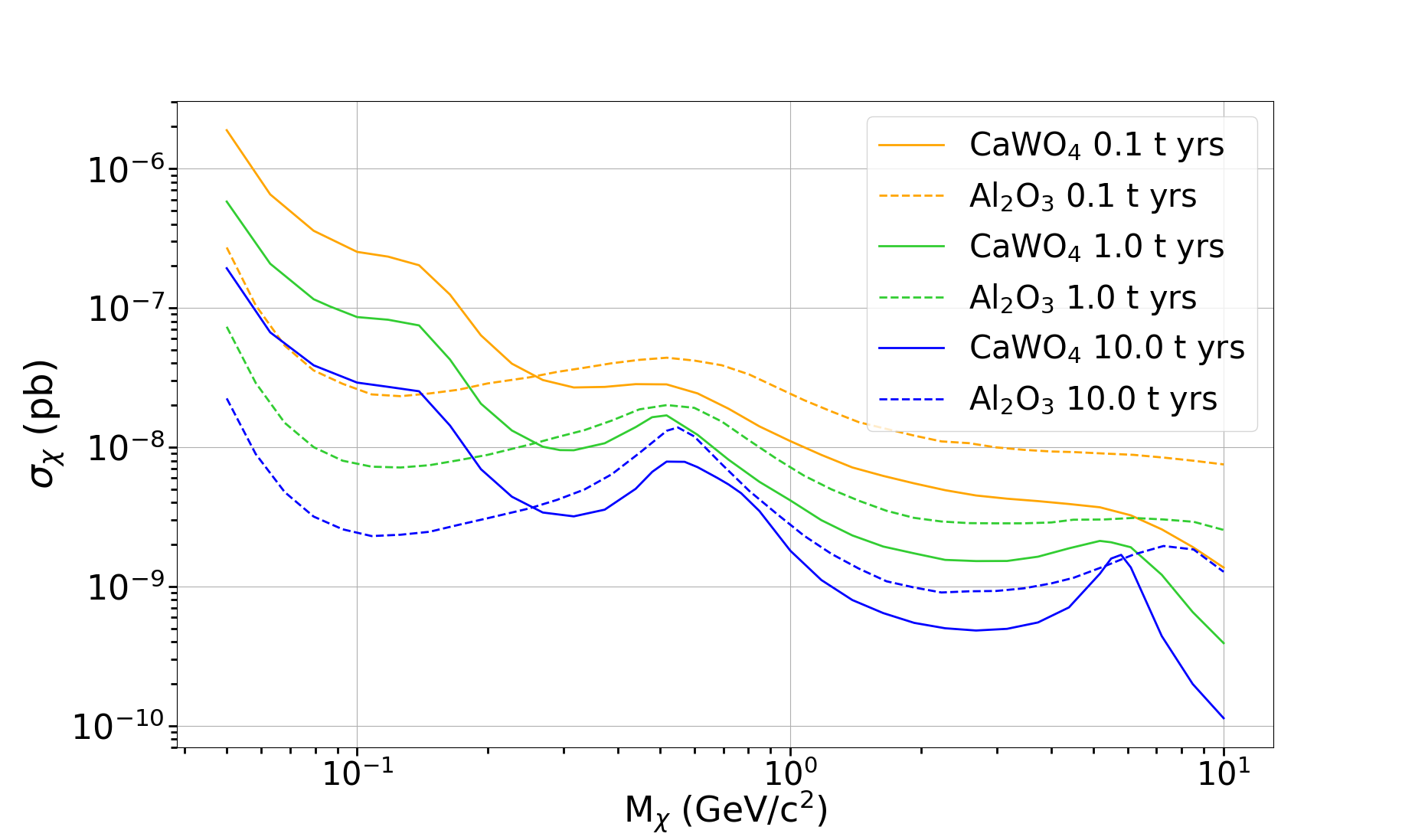}
	\caption{\label{fig:Both_SensLimits}Comparison of the 3$\sigma$ discovery limits for three different exposures and an energy threshold of 1$\,$eV for CaWO$_{4}$ detectors (solid lines) and Al$_{2}$O$_{3}$ detectors (dashed lines). Sapphire detectors have a much higher sensitivity at dark matter masses below about 250$\,$MeV/c$^{2}$. Below a dark matter mass of about 150$\,$MeV/c$^{2}$, the limit of a sapphire detector with an exposure of 0.1$\,$tonne years is comparable with the limit of a CaWO$_{4}$ detector with the same threshold, but two orders of magnitude higher exposure. At higher masses, CaWO$_{4}$ has a stronger sensitivity to a dark matter discovery.}
\end{figure}

A comparison of the discovery limits calculated for CaWO$_{4}$ and for Al$_{2}$O$_{3}$ is shown in Fig. \ref{fig:Both_SensLimits}. At masses below about 250$\,$MeV/c$^{2}$, the dark matter count rate in Al$_{2}$O$_{3}$ detectors surpasses the one in CaWO$_{4}$ detectors (with the same dark matter interaction cross section), while the number of neutrino events above the 1$\,$eV energy threshold is lower in the Al$_2$O$_{3}$ detectors. This leads to a much stronger dark matter discovery potential in sapphire detectors for masses below 250$\,$MeV/c$^{2}$. Below a dark matter mass of 150$\,$MeV/c$^{2}$ the limit of the Al$_{2}$O$_{3}$ detectors even reach the same level as the limit of a CaWO$_{4}$ detector with 100 times more exposure.

In conclusion, a lighter target, such as Al$_{2}$O$_{3}$, has a highly increased sensitivity at lower dark matter masses. On the other hand, sensitivity towards dark matter particles with higher masses is enhanced with a heavier target, such as CaWO$_{4}$. \\

\paragraph{Additional thoughts:} An additional factor that can help to distinguish between CE$\nu$NS and dark matter events that has not been included in this work is the expected seasonal modulation of the signals. Due to the relative motion of the Earth around the Sun, the rate of dark matter recoils is expected to follow an annual modulation \cite{RevModPhys.85.1561}. In a similar way, due to the changing distance of the Earth to the Sun between the Perihel and the Aphel, the rate of solar neutrino interactions is expected to be affected, scaling with the inverse of the distance squared. This leads to a difference of about 6.7$\,$\% with respect to the average rate. This modulation was confirmed in a measurement of the $^7$Be neutrino flux by Borexino \cite{AGOSTINI201721}. Including this time dependence into the neutrino and dark matter models in the likelihood formalism can provide additional discrimination power.

\section{Conclusion} \label{conclusion}

In this work we explore the sensitivities of cryogenic solid state detectors to a flavor independent measurement of the fluxes of solar neutrinos. We do this in the framework of low energy threshold $\mathcal{O}$(eV) and high exposure $\mathcal{O}$(tonne year) experiments. Furthermore, with many dark matter direct detection experiments approaching or already reaching the neutrino floor, it becomes increasingly important to understand the expected signals and investigate the impact on the search for dark matter.

\subsection{Sensitivity to solar neutrinos}

We focus on the necessary requirements for a measurement of the flux of solar neutrinos via CE$\nu$NS. $^{7}$Be and pp neutrinos are the dominant fluxes at low energies. We investigate the experimental settings needed to be able to correctly reconstruct these fluxes from the data. We find that the $^{7}$Be flux can be reconstructed with an uncertainty smaller than the theoretical one in an experiment using CaWO$_{4}$ detectors with an exposure of 1$\,$tonne$\,$years, a background rate smaller than 1/(keV$\,$kg$\,$d) and an energy threshold smaller than 6$\,$eV. In the case of an energy threshold of 1$\,$eV, an exposure of 0.1$\,$tonne$\,$years and a background rate smaller than 1/(keV$\,$kg$\,$d) are already sufficient. An experiment using Al$_{2}$O$_{3}$ detectors with an energy threshold of 1$\,$eV can reconstruct the pp neutrino flux with an uncertainty of less than 10$\,$\% in the case of an exposure of 0.5$\,$tonne$\,$years and a background rate of less than 1/(keV$\,$kg$\,$d).

In general, a low threshold and background level are more crucial than a high exposure in the case of pp and $^{7}$Be neutrinos. The precise reconstruction of the fluxes of these low energetic neutrinos in combination with the results of experiments measuring the electron scattering of such neutrinos opens up the possibility to test the MSW effect. \\

We additionally performed studies on the sensitivity to the summed flux of the CNO neutrinos. An accurate reconstruction of this flux in combination with the $^{7}$Be flux can help to contribute to the solution of the solar metallicity problem. We found that an exposure of 1$\,$tonne$\,$year with CaWO$_{4}$ detectors with energy thresholds below 6$\,$eV and a background rate of below $\mathcal{O}$(1)/(keV$\,$kg$\,$d) is necessary to be able to reject a low-metallicity (LZ) solar model with a significance of 3$\sigma$ at a 90$\,$\% C.L. At an exposure of 1$\,$tonne$\,$year with thresholds below 6$\,$eV and a background rate below 0.5/(keV$\,$kg$\,$d) the uncertainty on the reconstructed value of the CNO flux is below 20$\,$\% and thus lower than uncertainties from current experiments.

\subsection{Impact on dark matter searches}

Once the neutrino floor is reached, an accurate description of the expected spectral shape of the CE$\nu$NS of solar neutrinos in cryogenic detectors is necessary to disentangle them from potential dark matter events. In this work we use a likelihood framework to calculate discovery potentials for dark matter signals in the presence of a solar neutrino background for two different detector materials (CaWO$_{4}$, Al$_{2}$O$_{3}$). We show that a dark matter discovery is possible even below the classic definition of the neutrino floor in an experiment with an energy threshold of $\mathcal{O}$(eV) and an exposure of $\mathcal{O}$(tonne$\,$year). This is very important for experiments working on upgrades of their setups. We also show that lowering the threshold is more effective than increasing the exposure in order to drastically increase the sensitivity to light dark matter masses ($\leq \,\mathcal{O}$(100$\,$MeV/c$^{2}$)).

\begin{acknowledgments}

The authors would like to thank Lothar Oberauer for the helpful discussions and his support in the development of this work, and Heerak Banerjee for his invaluable help in outlining this work. This work has been partially funded by the DFG Collaborative Research Center 1258 "Neutrinos and Dark Matter".


\end{acknowledgments}

\appendix

\section{Dark Matter recoil rate}\label{DM_SolNu_models}

\begin{figure}[h]
\centering
\includegraphics[width=0.48\textwidth]{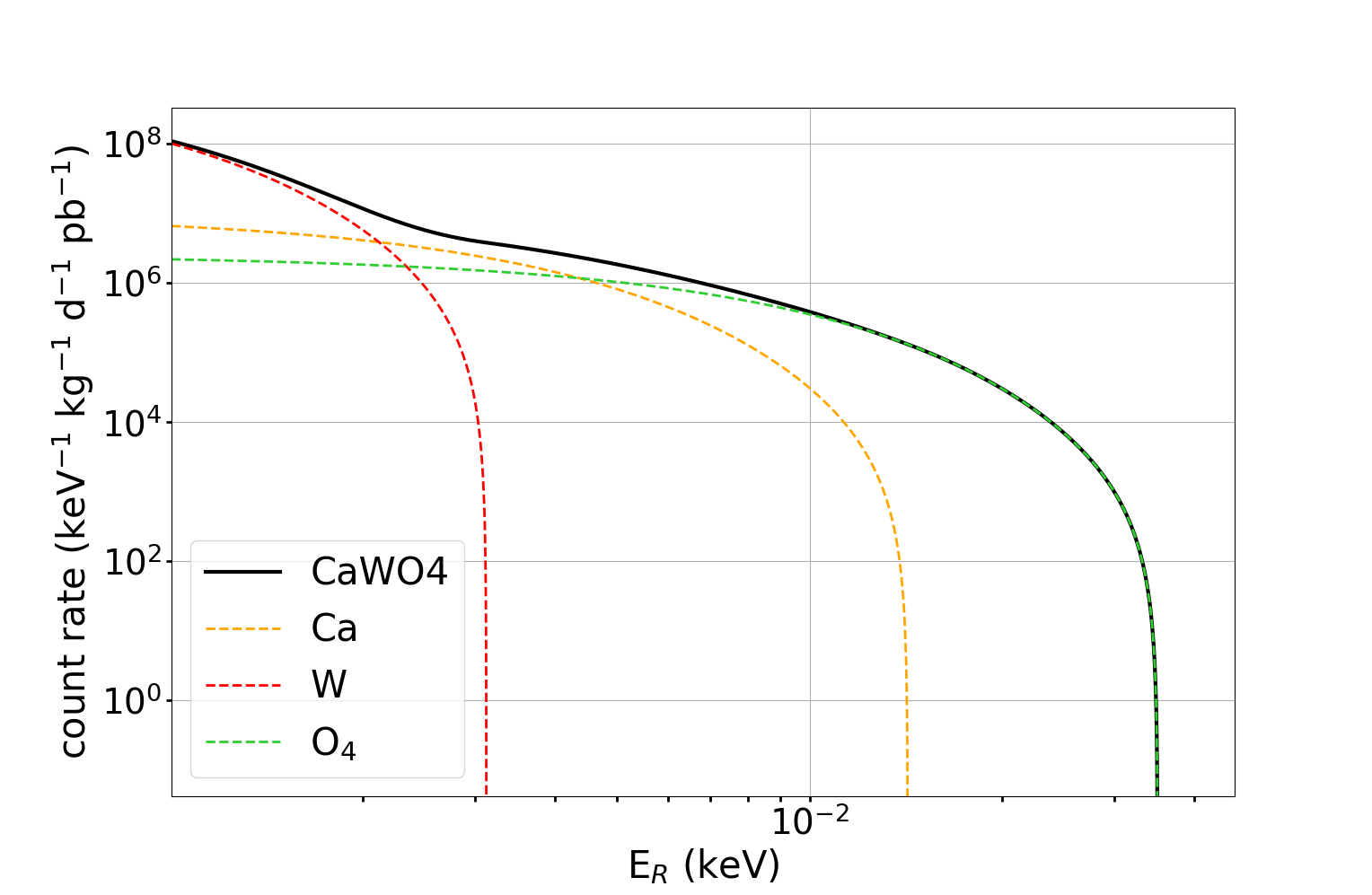}
\caption{\label{fig:recSpec}Expected differential recoil spectrum in a CaWO$_{4}$ detector for a 1$\,$GeV/c$^{2}$ dark matter particle, showing the contribution of the individual target nuclei to the recoil spectrum.}
\end{figure}

This section gives a short overview of the calculation of the event rates of spin-independent elastic dark matter-nucleus interactions in cryogenic solid state detectors. The differential recoil rate is expressed as:

\begin{equation} \label{eq:RecRate}
	\dfrac{\mathrm{d}R_{\chi}}{\mathrm{d}E_{\mathrm{R}}} \propto \dfrac{\rho_{\chi}}{m_{\chi}} \dfrac{m_{\mathrm{T}}}{2\mu_{\mathrm{n}}^{2}} A^{2} \sigma_{\mathrm{n}} F^{2}(q) \int\limits_{v_{\mathrm{min}}(E_{\mathrm{R}})}^{v_{\mathrm{esc}}} \dfrac{f(v)}{v} \, \mathrm{d}^{3}v 
\end{equation}

with $\rho_{\chi} \,$=$\,$0.3$\,$GeV/c$^{2}$/cm$^{3}$ being the standard local dark matter density, $m_{\chi}$ the mass of the dark matter particle, $m_{\mathrm{T}}$ the mass of the target nucleus and $\mu_{\mathrm{n}}$ the dark matter-nucleon reduced mass, $\mu_{\mathrm{n}} = m_{\chi}\cdot m_{\mathrm{n}}/(m_{\chi} + m_{\mathrm{n}})$. The material independent dark matter-nucleon interaction cross section $\sigma_{\mathrm{n}}$ is related to the spin-independent zero-momentum transfer cross section on a point-like nucleus, $\sigma_{0}$, via \cite{del2022theory}:

\begin{equation}
	\sigma_{0} = \sigma_{\mathrm{n}} \cdot A^{2} \cdot \dfrac{\mu_{\mathrm{T}}^{2}}{\mu_{\mathrm{n}}^{2}}
\end{equation}

introducing the $A^{2}$ (atomic number) into the recoil rate of eq. \ref{eq:RecRate}. To take the structure of the nucleus into account, the point-like cross section is multiplied with the nuclear form factor $F^{2}(q)$, depending on the momentum transfer $q = \sqrt{2m_{\mathrm{T}}E_{\mathrm{R}}}$, for which we use the Helm form factor parameterization \cite{PhysRev.104.1466}. The analytical description of the integral in eq. \ref{eq:RecRate} over the velocity distribution, assuming a non-rotating maxwellian model, is given in \cite{DONATO1998247}. The rotational velocity of the solar system is $v_{\mathrm{s}} \,$=$\,$220$\,$km/s and the the velocity of the rest frame of the Earth moving through the galaxy is $v_{0} \,$=$\,$232$\,$km/s \cite{DONATO1998247}. The lower boundary of the integral, $v_{\mathrm{min}}(E_{\mathrm{R}}) = \sqrt{\frac{m_{\mathrm{T}}E_{\mathrm{R}}}{2\mu_{\mathrm{T}}^{2}}}$, corresponds to the minimal velocity a dark matter particle needs to create a recoil of energy $E_{\mathrm{R}}$ in the target. The upper boundary is given by the galactic escape velocity $v_{\mathrm{esc}} \,$=$\,$544$\,$km/s \cite{10.1111/j.1365-2966.2007.11964.x}. An example of the expected recoil spectrum of a dark matter particle with a mass of $m_{\chi} \,$=$\,$1$\,$GeV/c$^{2}$ in a CaWO$_{4}$ detector for an exposure of 1$\,$kg$\,$d is shown in Fig. \ref{fig:recSpec}.

\bibliography{refs.bib}
\end{document}